\documentclass[english]{article}
\usepackage[T1]{fontenc}
\usepackage[utf8]{inputenc}
\usepackage{mathrsfs}
\usepackage{graphicx}
\usepackage{amsmath}
\usepackage{amssymb}
\usepackage{babel}
\begin{document}
\title{BLOCH MOTIONS AND SPINNING TOPS}
\author{Albert Huber\thanks{hubera@technikum-wien.at}\  \thanks{albert.huber@uni-graz.at},
Paul Schreivogl\thanks{schreivo@technikum-wien.at}}
\date{{\footnotesize UAS Technikum Wien - Faculty of Computer Science and
Applied Mathematics, Höchststädtplatz 6, 1200 Vienna, Austria, Karl-Franzens-University
Graz, Schubertstraße 51, Graz, Austria}}
\maketitle
\begin{abstract}
This work investigates the dynamics of closed quantum systems in the
Bloch vector representation using methods from rigid body dynamics
and the theory of integrable systems. To this end, equations of motion
for Bloch components are derived from the von Neumann equation that
are mathematically equivalent to equations of motion for a distribution
of point masses from classical mechanics. Furthermore, using the Heisenberg
equation, another system of Bloch vector equations is derived, which
constitutes an Euler-Poinsot system of the type commonly encountered
in the theory of torque-free spinning tops. This insight is used to
prove the Liouville integrability of the corresponding Hamilton equations
of motion and to construct an explicit closed-form solution using
the Arnold-Liouville theorem. Within the same framework, stability
criteria for quantum dynamics are then derived which correspond to
the Energy-Casimir method of classical Newtonian mechanics. Following
that, specific solutions to the equations of motion are constructed
that encode the complex dynamics of composite quantum systems. Eventually,
to show that this formalism provides concrete physical predictions,
an analogue of the intermediate axis theorem is derived and the effect
of oscillating entanglement is discussed. As a basis for this, special
types of solutions to the equations of motion are derived that constitute
oscillating entangled states, i.e., dynamical quantum states that
change their entanglement structure from maximally entangled to separable
and vice versa. 
\end{abstract}
{\footnotesize\textit{Key words: Quantum dynamics, Bloch vectors,
rigid body dynamics, integrable systems}}{\footnotesize\par}

\section*{Introduction }

It has long been known (and often been used) in quantum physics that
density operators characterizing the state of a quantum system can
be expressed in terms of different types of operator bases. Popular
choices leading to different types of decompositions of operators
are e.g. the standard projective basis, the polarization basis, the
generalized Gell-Mann basis and the Weyl operator basis built from
generalized Pauli operators such as clock and shift operators. But,
of course, there are many more. See, for example, \cite{bertlmann2008bloch}
for overviews.

Using the fact that not only density operators but all linear operators
in a Hilbert space can be decomposed in terms of the above basis,
equations of motion for Bloch vectors can be derived from the quantum
mechanical evolution equations, i.e., the Schrödinger, von Neumann,
or Dirac equations. Admittedly, the latter has long been known in
the literature, and specific solutions to the infamous Bloch and Maxwell-Bloch
equations have been found on various occasions, e.g., to describe
the effect of nuclear magnetic resonance and to derive Rabi oscillations
\cite{bloch1946nuclear,kittel2018introduction,torrey1956bloch}.

Despite this, as will be argued in this work, the equations of motion
for Bloch vectors and the geometric structure of the associated phase
space are still far from being fully explored. Moreover, many applications
of the formalism have not yet been exploited; particularly as regards
composite quantum systems, for which the equations of motion for the
Bloch components take on a rather complicated form.

To close this literature gap, equations of motion for Bloch components
are derived from the von Neumann equation in section one of this work,
which are mathematically equivalent to the equations of motion of
a distribution of point masses from classical theory of rigid body
dynamics. Furthermore, using the Heisenberg equation, a related set
of equations of motion is derived, which forms an Euler-Poinsot system,
as is commonly encountered in the theory of torque-free spinning tops.
To deduce these equations, an expansion of the von Neumann equation
is performed in the generalized Gell-Mann matrix basis (though it
is made clear that other types of operator bases could also be used
for the derivation). 

Following this calculation, it is shown in section two of this work
that the Hamiltonian dynamics of the problem prove to be Liouville
integrable and that a geometric approach to the subject can be adopted
for studying the resultant quantum dynamics using methods from rigid
body dynamics \cite{arnol2013mathematical,grammel1920kraftefreie,klein2008theory,klein2010theory,poinsot1851theorie}.
The latter leads to canonical variables in phase space and Hamiltonian
equations of motion for Bloch components, which are set up and studied
in detail. In this context, also novel types of stability criteria
are formulated, which constitute analogues to the Routh-Hurwitz criterion
and nonlinear criteria following from the Energy-Casimir method of
the classical theory of dynamical systems \cite{hurwitz1963bedingungen,routh1877treatise}
(though only the latter prove relevant for unitary dynamics). To identify
all constants of motion of the deduced Euler-Poinsot system, it is
further shown that the deduced Hamiltonian dynamics can be traced
back in the qubit case to the Neumann model of classical Hamiltonian
mechanics and the Hamiltonian version of the Euler-Poinsot model.
The latter is used to construct operator-valued constants of motion
using the Lax formulation of the theory, where the mentioned quantities
constitute analogues of the Uhlenbeck first integrals of motion in
the qubit case \cite{babelon2003introduction} and Lax-type conserved
quantities \`a la Sklyanin and Hitchin in the higher-dimensional case
\cite{babelon2003introduction,hitchin1987stable,kulish1979quantum,sklyanin1989separation}.
The latter are derived in the higher-dimensional context by employing
the so-called Adler-Kostant-Symes scheme using a Gaudin-type Lax matrix
ansatz, whereby an explicit closed-form solution of the Bloch vector
dynamics in the form of theta functions is derived based on the Arnold-Liouville
theorem \cite{arnol2013mathematical,babelon2003introduction,landau1982mechanics}
using the Its-Matveev formula for the Baker-Akhiezer function \cite{its1975schrodinger,krichever1977methods}.
The solution presented, it seems, is new and not yet known in the
literature.

After obtaining these results, in section three of the paper, the
developed formalism is applied to the more complicated case of composite
quantum systems. Specifically, using again an operator basis expansion
of the von Neumann equation, but now in terms of tensor products of
generalized Gell-Mann matrices, Bloch component equations are formulated
for the multipartite case, i.e. special solutions of the equations
of motion are constructed that specify the dynamics of composite quantum
systems. To the best of our knowledge, neither these equations for
the Bloch components nor the underlying Bloch-type Euler-Poinsot equations
have yet been studied or even presented in the literature. The resultant
composite dynamics are quite complex, as demonstrated towards the
end of section three of the work by the concrete example of Heisenberg
dimer system coupled to a small magnetic field. Also, it is clarified
that the quantum dynamics of product states is less difficult to handle
than that of separable and, in particular, entangled quantum states. 

Ultimately, in order to show that the formalism developed allows for
concrete physical predictions, a generalized higher-dimensional Bloch
analogue of the intermediate axis theorem (alternatively referred
to as tennis racket theorem or Dzhanibekov-Poinsot effect) is derived
and the effect of oscillating entanglement is discussed in the fourth
and final section of this work. As a basis for the latter, solutions
of the Bloch equations are identified that constitute oscillating
entangled states, i.e. dynamical quantum states that change their
entanglement structure over time from separable to maximally entangled
and vice versa. 

The paper concludes with a summary of key results and an outlook on
possible lines of future research.

\section{Bloch Motions in Isolated Quantum Systems I: The Equations of Motion }

In a closed quantum system, the unitary evolution of a (qudit) state
$\rho\in\mathcal{B}(\mathcal{H})$ under a Hamiltonian $H(t)$ on
a complex Hilbert space $\mathcal{H}$ is described by the von Neumann
equation 
\begin{equation}
\dot{\rho}=-i[H,\rho],
\end{equation}
where in this context (as throughout this paper), for simplicity,
$\hbar$ is set to one. Given that $\rho_{0}$ is some initial state
and $U(t,t_{0})$ is a unitary operator meeting the initial condition
$U(t,t_{0})=\textbf{1}_{d}$, the solution of this equation is known
to be the density matrix 
\begin{equation}
\rho(t)=U(t,t_{0})\rho_{0}U^{\dagger}(t,t_{0}).
\end{equation}
For the special case of a time-independent Hamiltonian $H_{0}$, the
solution of $(1)$ is simply $\rho(t)=e^{-i(t-t_{0})H_{0}}\rho_{0}e^{i(t-t_{0})H_{0}}$
as $U(t,t_{0})=\exp\{-i(t-t_{0})H_{0}\}$ applies in this particular
context. Alternatively, in case of a time-dependent Hamiltonian, solution
$(2)$ is given by an infinite Volterra series, i.e. an infinite time-ordered
product of the form $U(t,t_{0})=\mathcal{T}\exp\{-i\overset{t}{\underset{t_{0}}{\int}}H(\tau)d\tau\}$. 

Using the fact that any operator can be expanded with respect to a
corresponding type of operator basis, it becomes clear that both the
density operator and the Hamiltonian can be expanded in the generalized
Gell-Mann basis $\{\Lambda_{\alpha}\}$, i.e.

\begin{equation}
\rho=\frac{1}{d}\textbf{1}_{d}+\vec{\varrho}\cdot\vec{\Lambda},\;H=h\textbf{1}_{d}+\vec{h}\cdot\vec{\Lambda},
\end{equation}
where the generalized Gell-Mann matrices take the form $\Lambda^{S}_{jk}=\sqrt{\frac{d}{2}}(\vert k\rangle\langle j\vert+\vert j\rangle\langle k\vert)$,
$\Lambda^{A}_{jk}=-i\sqrt{\frac{d}{2}}(\vert k\rangle\langle j\vert-\vert j\rangle\langle k\vert)$
and $\Lambda^{D}_{k}=\sqrt{\frac{d}{k(k+1)}}(\overset{k}{\underset{j=1}{\sum}}\vert j\rangle\langle j\vert-k\vert k+1\rangle\langle k+1\vert)$
and the shorthand notation $\vec{\xi}\vec{\Lambda}=\overset{N-1}{\underset{k=1}{\sum}}\xi_{k}\Lambda_{k}$
with $N=d^{2}$ is used; see here e.g. \cite{bertlmann2008bloch,loubenets2021bloch}
for further details. 

The generalized Gell-Mann matrices, same as the 'conventional' Gell-Mann
matrices, are subject to the commutation relations

\begin{align}
\{\Lambda_{i},\Lambda_{j}\} & =\frac{4}{d}\delta_{ij}\textbf{1}_{d}+2\overset{N-1}{\underset{k=1}{\sum}}g_{ijk}\Lambda_{k},\\{}
[\Lambda_{i},\Lambda_{j}] & =2i\overset{N-1}{\underset{k=1}{\sum}}f_{ijk}\Lambda_{k}\nonumber 
\end{align}
and therefore meet the condition

\begin{equation}
\Lambda_{i}\Lambda_{j}=\frac{2}{d}\delta_{ij}\textbf{1}_{d}+\overset{N-1}{\underset{k=1}{\sum}}[g_{ijk}+if_{ijk}]\Lambda_{k}.
\end{equation}
Expanding the density operator and the Hamiltonian as described above,
it becomes clear that the Bloch coefficients $\vec{\rho}$, $\vec{h}$
and $h$ encode the entire dynamics of the theory, since only they
are time-dependent. Based on this observation, the von Neumann equation
$(1)$ can be written as

\begin{equation}
\dot{\varrho}_{k}=\overset{N-1}{\underset{l=1}{\sum}}\mathcal{J}_{kl}\varrho_{l}
\end{equation}
in Bloch coefficients, where $\mathcal{J}_{kl}=2\overset{N-1}{\underset{m=1}{\sum}}f_{klm}h_{m}$
applies by definition. Abandoning the used index notation, this evolution
equation can be rewritten in the simpler form
\begin{equation}
\dot{\vec{\varrho}}=\mathcal{J}\vec{\varrho},
\end{equation}
where $\mathcal{J}(t)$ is a skew-symmetric matrix. 

In the qubit case, the meaning of this equation becomes readily apparent.
For, in the mentioned case, the Bloch decompositions $(3)$ take the
simple form
\begin{equation}
\rho=\frac{1}{2}\textbf{1}_{2}+\vec{\varrho}\cdot\vec{\sigma},\;H=h\textbf{1}_{2}+\vec{h}\cdot\vec{\sigma},
\end{equation}
and the structure constants $f_{jkl}$ of the Lie algebra coincide
with the Levi-Civita symbol $\epsilon_{jkl}$. As a result, using
the definition $\vec{\omega}:=2\vec{h}$, it is found that relation
$(7)$ takes the simple form

\begin{equation}
\dot{\vec{\varrho}}=\mathcal{J}\vec{\varrho}=\vec{\omega}\times\vec{\varrho},
\end{equation}
where $\mathcal{J}(t)$ is a time-dependent matrix. Therefore, one
thus recovers the familiar result that the Bloch vector $\vec{\rho}$
rotates by the action of a unitary transformation $U(t,t_{0})$ on
$\rho_{0}$ along a spherical shell within the Bloch sphere. In the
higher-dimensional case, it can therefore be expected that the Bloch
vector undergoes some type of generalized rotation on some fixed orbit
within the generalized Bloch multisphere.

By taking the second time derivative and using once more the von Neumann
equation, the double commutator relation

\begin{equation}
\ddot{\rho}=-i[\dot{H},\rho]-[H,[H,\rho]]
\end{equation}
can be derived. The latter can be used to derive the equation of motion 

\begin{align}
\ddot{\varrho}_{j} & =2\overset{N-1}{\underset{k,l=1}{\sum}}[f_{jkl}\dot{h}_{k}\varrho_{l}+f_{jkl}h_{k}\dot{\varrho}_{l}]=\\
 & =2\overset{N-1}{\underset{k,l=1}{\sum}}[f_{jkl}\dot{h}_{k}\varrho_{l}+2f_{jkl}h_{k}\overset{N-1}{\underset{m,n=1}{\sum}}f_{lmn}h_{m}\varrho_{n}],\nonumber 
\end{align}
for the Bloch vector, where the latter equation can be rewritten more
compactly in the form

\begin{equation}
\ddot{\varrho}_{j}=\overset{N-1}{\underset{n=1}{\sum}}\mathcal{K}_{jn}\varrho_{n},
\end{equation}
with $\mathcal{K}_{jn}=2\overset{N-1}{\underset{k=1}{\sum}}f_{jkn}[\dot{h}_{k}+2\overset{N-1}{\underset{m,l=1}{\sum}}f_{jml}h_{m}h_{k}]$.
Without index notation, this result can be recast in the even simpler
form

\begin{equation}
\ddot{\vec{\varrho}}=\mathcal{K}\vec{\varrho},
\end{equation}
where $\mathcal{K}=\dot{\mathcal{J}}+\mathcal{J}^{2}$ applies in
the given context. 

In the qubit case, equation $(13)$ takes a much simpler form, which
is
\begin{equation}
\ddot{\vec{\varrho}}=\dot{\vec{\omega}}\times\vec{\varrho}+\vec{\omega}\times\dot{\vec{\varrho}}=\dot{\vec{\omega}}\times\vec{\varrho}+\vec{\omega}\times(\vec{\omega}\times\vec{\varrho}).
\end{equation}
To interpret this result, a comparison with classical Newtonian dynamics
seems appropriate. This is because the derived expression contains
two terms, one of which is mathematically identical to an Euler acceleration
term and the other to a centrifugal acceleration term from the theory
of classical mechanics. The derived equations, however, originate
from the von Neumann equation and therefore fully characterize the
dynamics of closed quantum systems on the level of Bloch vectors of
qubits. 

From a purely mathematical point of view, though, the derived equations
match the equations of motion for a rotating point particle from classical
Newtonian dynamics. Furthermore, given the case that the density operator
considered in $(2)$ and $(3)$ is a mixed state of the form $\rho(t)=\overset{\mu}{\underset{k=1}{\sum}}\lambda_{\nu}\rho_{\nu}(t)$,
where the density operators $\rho_{\nu}(t)$ with $\nu\in\{1,2,...,\mu\}$
are pure states meeting the condition $\rho^{2}_{\nu}=\rho_{\nu}$,
a decomposition of the form $(3)$ yields $\vec{\rho}(t)=\overset{\mu}{\underset{\nu=1}{\sum}}\lambda_{\nu}\vec{\rho}_{\nu}(t)$
for the Bloch vector corresponding to $\rho(t)$. The deduced vector
relation, however, again exactly resembles a relation from classical
mechanics, namely that for the center-of-mass vector $\vec{\rho}(t)$
of a discrete mass distribution ${\lambda_{1},\lambda_{2},...,\lambda_{\mu}}$,
i.e., a distribution of point masses with position vectors ${\vec{\rho}_{1}(t),\vec{\rho}_{2}(t),...,\vec{\rho}_{\mu}(t)}$,
all of which have the same length $\mathcal{R}^{2}=\overset{\mu}{\underset{\nu=1}{\sum}}\vec{\rho}^{2}_{\nu}$. 

As explained below, this analogy can be extended even further, in
the sense that any type of quantum dynamics as described above can
always be assigned an inertia tensor $\mathcal{I}_{jk}$, and thus
relevant properties of a rigid body. In the case of a pure state,
as it turns out, this inertia tensor is just that of a single point
mass, while in the case of a mixed state, it is that of a distribution
of point masses. This inertia tensor will be referred to alternately
as \textit{associated inertia tensor} or \textit{Bloch-type inertia
tensor} in the following. But, as already mentioned, further details
will only be given below.

Before that, however, let it first be noted that, at first glance,
something appears to go wrong with the above analogy to Newtonian
physics: equation $(14)$ contains no Coriolis force term. However,
as a deeper comparison shows, there is nothing wrong with the analogy,
as the absence of the term can be plausibly explained. To see this,
it proves convenient to switch again for a moment to classical Newtonian
physics and consider an inertial frame $\mathcal{O}$ and a rotating
reference frame $\mathcal{O}'$ with corresponding position vectors
that are related by the transformations $\vec{\varrho}(t)=R(t)\vec{\varrho}\,'$
and $\vec{\varrho}\,'=R^{\intercal}(t)\vec{\varrho}(t)$, where $R(t)$
is some orthogonal matrix. For example, in this context, it could
be assumed that the matrix $R(t)$ is a suitable non-constant rotation
matrix with generic time-dependent angles. Assuming then that $\vec{\varrho}\,'$
is at rest in $\mathcal{O}'$, the equations of motion 

\begin{equation}
\dot{\vec{\rho}}\,'=\dot{\vec{\varrho}}-\vec{\omega}\times\vec{\varrho}=0
\end{equation}
and
\begin{equation}
\ddot{\vec{\rho}}\,'=\ddot{\vec{\rho}}-\dot{\vec{\omega}}\times\vec{\varrho}-\vec{\omega}\times(\vec{\omega}\times\vec{\varrho})=0
\end{equation}
can be deduced. This makes it clear, however, that the equations of
motion $(9)$ and $(14)$ for the qubit Bloch vector components are
identical to the equations of motion of a point mass (respectively,
the center-of-mass of a distribution of point masses), whereby the
latter rotates around the origin of the considered inertial frame
while being fully at rest relative to the rotating reference frame;
a special case that can occur quite naturally in Newtonian mechanics.
The derived equations of motion $(9)$ and $(14)$ for the qubit Bloch
vector are thus mathematically equivalent to the equations of motion
$(15)$ and $(16)$ from classical Newtonian dynamics.

Note that relation $(14)$ can be simplified using the identity $\vec{\omega}\times(\vec{\omega}\times\vec{\varrho})=(\vec{\omega}\vec{\varrho})\vec{\omega}-\vec{\omega}^{2}\vec{\varrho}$,
which may ultimately be used to derive the result
\begin{equation}
\mathcal{K}_{ij}=\omega_{i}\omega_{j}-\omega^{2}\delta_{ij},
\end{equation}
where $\omega^{2}:=4\vec{h}^{2}$ applies by definition. In case that
$\dot{\vec{\omega}}=0$ applies the solution of equation $(14)$ reads

\begin{equation}
\vec{\varrho}(t)=\vec{\varrho}_{01}e^{i\mathfrak{K}}+\vec{\varrho}_{02}e^{-i\mathfrak{K}}=\vec{A}\cos\mathfrak{K}+\vec{B}\sin\mathfrak{K},
\end{equation}
where $\mathfrak{K}(t)=\overset{t}{\underset{0}{\int}}\sqrt{\mathcal{\mathcal{K}}}d\tau$
applies by definition and the vectors $\vec{\rho}_{0j}$ resp. $\vec{A}$,
$\vec{B}$ are constant vectors encoding information about the orientation
of the axis of rotation of the system. 

That said, it may be noted that $\mathcal{\mathcal{K}}$ can be diagonalized
so that $\mathcal{K}_{ij}=\mathcal{K}_{i}\delta_{ij}$ applies, which
simplifies the calculation of $\mathfrak{K}(t)$. In case that $\dot{\vec{\omega}}=0$
and all eigenvalues of $\mathcal{\mathcal{K}}$ are equal to $\mathcal{K}_{i}=-\omega^{2}=const.$,
a case in which obviously the conditions $\vec{\omega}\vec{\varrho}=0$
and $[\mathcal{K}(t_{i})\mathcal{K}(t_{j})]=0$ are satisfied for
all $i,j$, the equations of motion become rather simple. For in the
mentioned case, which corresponds to a popular text book model of
Newtonian dynamics, the Bloch vector performs simply a stationary
rotation around a single axis on a spherical orbit within the Bloch
sphere. The latter is governed by the equation of motion for the spherical
harmonic oscillator reading
\begin{equation}
\ddot{\vec{\varrho}}=-\omega^{2}\vec{\varrho},
\end{equation}
which can readily be solved. That is, solution $(18)$ takes the comparatively
simple form $\vec{\varrho}(t)=\vec{\rho}_{01}e^{i\omega t}+\vec{\rho}_{02}e^{-i\omega t}=\vec{A}\cos\omega t+\vec{B}\sin\omega t$.

The latter proves of interest not least because there is a linear
operator
\begin{equation}
\mathcal{C}=\frac{1}{2i}[\dot{\rho},\rho],
\end{equation}
which constitutes a constant of motion in case that the Bloch equations
$(13)$ reduce to the special case given by $(19)$, as can readily
be checked using the Heisenberg equations of motion

\begin{equation}
\dot{X}=i[H,X]+\frac{\partial X}{\partial t}.
\end{equation}
For in the mentioned case, given that $\varrho=\rho-\frac{1}{d}\textbf{1}_{d}=\vec{\varrho}\cdot\vec{\Lambda}$,
it thus becomes clear that the double commutator relation $(10)$
can be rewritten in the maximally simple form 
\begin{equation}
\ddot{\rho}=\ddot{\varrho}=-\omega^{2}\varrho,
\end{equation}
thereby showing that $\dot{\mathcal{C}}=i[H,\mathcal{C}]+\frac{\partial C}{\partial t}=0$.
Using the decomposition
\begin{equation}
\mathcal{C}=\vec{\ell}\cdot\vec{\Lambda},
\end{equation}
which reduces to the form $\mathcal{C}=\vec{\ell}\cdot\vec{\sigma}$
in the qubit case, this result can readily be generalized. To show
this, it may be taken into account that the components of the Bloch
vector $\vec{\ell}$ read

\begin{equation}
\ell_{i}=\overset{N-1}{\underset{j,k=1}{\sum}}f_{ijk}\rho_{j}\dot{\rho}_{k},
\end{equation}
thereby taking the particularly simple form

\begin{equation}
\vec{\ell}=\mathcal{I}\vec{\omega}=\vec{\varrho}\times\dot{\vec{\varrho}}
\end{equation}
in the qubit case. Note that the matrix $\mathcal{I}$ corresponds
to the above-mentioned Bloch-type inertia tensor
\begin{equation}
\mathcal{I}_{jk}=\overset{N-1}{\underset{l,m,n=1}{\sum}}f_{jlm}f_{kmn}\rho_{l}\rho_{n},
\end{equation}
which can be written as
\begin{equation}
\mathcal{I}_{jk}=\mathcal{R}^{2}\delta_{jk}-\rho_{j}\rho_{k},
\end{equation}
in the qubit case, where the radial parameter $0\leq\mathcal{R}\leq1$
denotes the length of the Bloch vector, i.e. $\mathcal{R}=\sqrt{\vec{\varrho}\vec{\varrho}}$.
This associated inertia tensor agrees with that of a rotating Newtonian
(unit mass) point particle, which, however, makes it clear that the
definition of the considered angular momentum expression $(25)$ and
all equations of motion containing this expression are mathematically
equivalent to those of the Newtonian counterpart. But even more than
that: In case that a mixed state $\rho(t)=\overset{\mu}{\underset{\nu=1}{\sum}}\lambda_{\nu}\rho_{\nu}(t)$
with Bloch vector $\vec{\rho}(t)=\overset{\mu}{\underset{\nu=1}{\sum}}\lambda_{\nu}\vec{\rho}_{\nu}(t)$
is considered, where the latter evidently looks again like a center-of-mass
vector of Newtonian mechanics for a distribution of point masses ${\lambda_{1},\lambda_{2},...,\lambda_{\mu}}$
with positions ${\vec{\rho}_{1}(t),\vec{\rho}_{2}(t),...,\vec{\rho}_{\mu}(t)}$,
the Bloch-type inertia tensor takes exactly the form one would expect
in the qubit case, which is
\begin{equation}
\mathcal{I}_{jk}=\overset{\mu}{\underset{\nu=1}{\sum}}\lambda_{\nu}[\mathcal{R}^{2}_{\nu}\delta_{jk}-\rho_{j\nu}\rho_{k\nu}].
\end{equation}
Note that the $SU(d)$-analog of this expression reads 
\begin{equation}
\mathcal{I}_{jk}=\overset{\mu}{\underset{\nu=1}{\sum}}\lambda_{\nu}\overset{N-1}{\underset{i,l,m=1}{\sum}}f_{ijl}f_{lkm}\rho_{i\nu}\rho_{m\nu}
\end{equation}
with continuum limit
\begin{equation}
\mathcal{I}_{jk}=\int d^{N-1}\rho\,\chi(\rho)\overset{N-1}{\underset{i,l,m=1}{\sum}}f_{ijl}f_{lkm}\rho_{i}\rho_{m},
\end{equation}
where $\chi(\rho)$ is a probability distribution with mass density
as its Newtonian analogue. As may also be noted, expression $(29)$
results from $(30)$ in case that an ansatz of the form $\chi(\rho)=\overset{\mu}{\underset{\nu=1}{\sum}}\lambda_{\nu}\delta^{(N-1)}(\vec{\rho}-\vec{\rho}_{\nu})$
is made in this context, whereby the qubit case $(28)$ arises for
$N=d^{2}=4$, i.e. $\chi(\rho)=\overset{\mu}{\underset{\nu=1}{\sum}}\lambda_{\nu}\delta^{(3)}(\vec{\rho}-\vec{\rho}_{\nu})$.
It is an interesting question whether alternative examples for $\chi(\rho)$
can be found, which lead to an inertia tensor that differs from $(29)$.

In any case, after clarifying the above, let it next be noted that
the Heisenberg equation
\begin{equation}
\dot{\mathcal{C}}=\frac{\partial C}{\partial t}+i[H,\mathcal{C}]=0
\end{equation}
can be recast in the form

\begin{equation}
\frac{\partial\ell_{j}}{\partial t}-\overset{N-1}{\underset{k=1}{\sum}}\mathcal{J}_{jk}\ell_{k}=0
\end{equation}
when being expanded in a generalized Gell-Mann matrix basis, where
the latter equation can be written as

\begin{equation}
\frac{\partial\vec{\ell}}{\partial t}-\vec{\omega}\times\vec{\ell}=0
\end{equation}
in the qubit case. By using this result in combination with $(26)$
and $(28)$ and by performing a principal axis transformation of $(27)$
resp. $(29)$ such that $\mathcal{I}_{jk}=\mathcal{I}_{j}\delta_{jk}$
with $\dot{\mathcal{I}}_{j}=0$ applies in this context, it is found
that relation $(32)$ can be further converted to give the system
of equations
\begin{equation}
\mathcal{I}_{j}\frac{\partial\omega_{j}}{\partial t}-\overset{N-1}{\underset{k,l=1}{\sum}}(\mathcal{I}_{k}-\mathcal{I}_{l})f_{jkl}\omega_{k}\omega_{l}=0.
\end{equation}
This system of equations takes the form
\begin{align}
\mathcal{I}_{1}\frac{\partial\omega_{1}}{\partial t}-(\mathcal{I}_{2}-\mathcal{I}_{3})\omega_{2}\omega_{3} & =0,\\
\mathcal{I}_{2}\frac{\partial\omega_{2}}{\partial t}-(\mathcal{I}_{3}-\mathcal{I}_{1})\omega_{1}\omega_{3} & =0,\nonumber \\
\mathcal{I}_{3}\frac{\partial\omega_{3}}{\partial t}-(\mathcal{I}_{1}-\mathcal{I}_{2})\omega_{1}\omega_{2} & =0,\nonumber 
\end{align}
in the qubit case $(34)$ and thus can be identified in the spirit
of classical Newtonian mechanics as an Euler-Poinsot system. However,
this makes it clear that solutions to the Euler-Poinsot problem of
rigid body dynamics give rise to a special type of solutions of equations
$(34)$ and that a classic Poinsot construction can be used to analyze
geometrically the corresponding Bloch motions, i.e. the motions of
the Bloch vector (of both the density operator and the Hamiltonian)
within the Bloch sphere. 

Before proceeding, a few comments are in order. As regards equations
$(34)$ and $(35)$, the question naturally arises as to why considering
these equations should be of any interest at all. To see why, let
it be noted that the stability of quantum dynamics can only be guaranteed
for density operators and the Hamiltonians with suitable types of
Bloch vectors. That is to say, unless one Bloch vector is properly
matched to the other, the dynamical behavior of the quantum system
may become unstable under linear perturbations, similarly to what
occurs in classical systems. Consequently, both Bloch vectors need
to be appropriately adjusted towards each other to avoid dynamical
instability. One method for determining restrictions on the corresponding
Bloch vectors is to demand the nonlinear stability of equation $(33)$,
which reduces to $(35)$ in the qubit case. The key idea here is that
the stability of the dynamics at the level of Bloch vectors also induces
the stability of the dynamics at the operator level. The idea of imposing
both conditions is based on the following observation: The von Neumann
equation $(1)$ not only allows the density operator to be determined
with respect to a given Hamiltonian, but also, conversely, the Hamiltonian
to be determined with respect to a given density operator. These two
ways of reading the equation show that, depending on the perspective,
different types of dynamical instabilities can arise at the Bloch
vector level, namely linear and nonlinear instabilities, both of which
should be excluded in order to guarantee the stability of the quantum
dynamics. This point will be discussed in much greater detail below
and for the higher-dimensional case in the second section of this
paper. 

That said, it appears a good strategy to first study $(35)$ extensively,
and only thereafter to generalize the results to the higher-dimensional
case. As a first step for doing so, it may be noted that the system
$(35)$ is fully integrable in cases where the Bloch-type inertia
tensor differs from that of a point particle, i.e. $\mathcal{I}_{1}\neq\mathcal{I}_{2}\neq\mathcal{I}_{3}$,
as there are two first integrals of motion \cite{babelon2003introduction,murakami2021analytical}
\begin{align}
\mathcal{H} & =\frac{1}{2}\overset{3}{\underset{k=1}{\sum}}\ell_{k}\omega_{k}=\frac{1}{2}\overset{3}{\underset{k=1}{\sum}}\mathcal{I}_{k}\omega^{2}_{k},\\
\ell^{2} & =\overset{3}{\underset{k=1}{\sum}}\ell^{2}_{k}=\overset{3}{\underset{k=1}{\sum}}\mathcal{I}^{2}_{k}\omega^{2}_{k}.\nonumber 
\end{align}
The existence of these constants allows the definition of the further
constants of motion $\varsigma:=\frac{2\mathcal{H}}{\ell^{2}}$ and
$\varpi:=\varsigma^{-1}=\frac{\ell^{2}}{2\mathcal{H}}$, both of which
will play a role for writing down solutions to $(33)$ in the following.

As a basis for the latter, let the two fully asymmetric cases be distinguished
in which $\mathcal{I}_{1}<\varpi<\mathcal{I}_{2}<\mathcal{I}_{3}$
resp. $\mathcal{I}_{1}<\mathcal{I}_{2}<\varpi<\mathcal{I}_{3}$ applies.
In the first case $\varpi<\mathcal{I}_{2}$ , the solutions of $(33)$
read
\begin{align}
\omega_{1}(t) & =\Gamma_{1}dn(n_{1}t,k_{1}),\\
\omega_{2}(t) & =\Gamma_{2}sn(n_{1}t,k_{1}),\nonumber \\
\omega_{3}(t) & =\Gamma_{3}cn(n_{1}t,k_{1}),\nonumber 
\end{align}
whereas in the second case $\mathcal{I}_{2}<\varpi$ they read
\begin{align}
\omega_{1}(t) & =\Gamma_{1}cn(n_{2}t,k_{2}),\\
\omega_{2}(t) & =\Gamma_{2}sn(n_{2}t,k_{2}),\nonumber \\
\omega_{3}(t) & =\Gamma_{3}dn(n_{2}t,k_{2}),\nonumber 
\end{align}
provided that $dn(n_{i}t,k_{i})$, $sn(n_{i}t,k_{i})$ and $cn(n_{i}t,k_{i})$
with $i=1,2$ are Jacobi elliptic functions and the introduced constants
take the form $\Gamma_{1}=\ell\sqrt{\frac{1-\varsigma\mathcal{I}_{3}}{\mathcal{I}^{2}_{1}-\mathcal{I}_{1}\mathcal{I}_{3}}}$,
$\Gamma_{2}=\ell\sqrt{\frac{1-\varsigma\mathcal{I}_{3}}{\mathcal{I}^{2}_{2}-\mathcal{I}_{2}\mathcal{I}_{3}}}$,
$\Gamma_{3}=\ell\sqrt{\frac{1-\varsigma\mathcal{I}_{1}}{\mathcal{I}^{2}_{3}-\mathcal{I}_{1}\mathcal{I}_{3}}}$
and $n_{1}=\ell\sqrt{\frac{(\mathcal{I}_{2}-\mathcal{I}_{3})(\varsigma\mathcal{I}_{1}-1)}{\mathcal{I}_{1}\mathcal{I}_{2}\mathcal{I}_{3}}}$,
$k_{1}=\sqrt{\frac{\mathcal{I}_{1}-\mathcal{I}_{2}}{\mathcal{I}_{2}-\mathcal{I}_{3}}\frac{1-\varsigma\mathcal{I}_{3}}{\varsigma\mathcal{I}_{1}-1}}$
resp. $n_{2}=\ell\sqrt{\frac{(\mathcal{I}_{1}-\mathcal{I}_{2})(1-\varsigma\mathcal{I}_{3})}{\mathcal{I}_{1}\mathcal{I}_{2}\mathcal{I}_{3}}}$,
$k_{2}=\sqrt{\frac{\mathcal{I}_{2}-\mathcal{I}_{3}}{\mathcal{I}_{1}-\mathcal{I}_{2}}\frac{1-\varsigma\mathcal{I}_{1}}{\varsigma\mathcal{I}_{3}-1}}$.
Since the solutions thus obtained are simply those of the torque-free
asymmetric Euler top of rigid body dynamics, the curves generated
by those solutions are epicycloidal and pericycloidal polhodes lying
on the Poinsot ellipsoid, which are divided by the so-called separating
polhodes or separatrix. The polhodes are closed spatial curves that
are reflected on each of the three principal planes.

Using $(36)$, the generic relations
\begin{align}
\mathcal{I}_{2}(\mathcal{I}_{2}-\mathcal{I}_{1})\omega^{2}_{2}+\mathcal{I}_{3}(\mathcal{I}_{3}-\mathcal{I}_{2})\omega^{2}_{3} & =\ell^{2}-2\mathcal{H}\mathcal{I}_{1}\\
\mathcal{I}_{1}(\mathcal{I}_{2}-\mathcal{I}_{1})\omega^{2}_{1}+\mathcal{I}_{3}(\mathcal{I}_{2}-\mathcal{I}_{3})\omega^{2}_{3} & =2\mathcal{H}\mathcal{I}_{2}-\ell^{2}\nonumber \\
\mathcal{I}_{1}(\mathcal{I}_{3}-\mathcal{I}_{1})\omega^{2}_{1}+\mathcal{I}_{2}(\mathcal{I}_{3}-\mathcal{I}_{2})\omega^{2}_{2} & =2\mathcal{H}\mathcal{I}_{3}-\ell^{2}\nonumber 
\end{align}
can be derived, whereby the first and third equations only have positive
coefficients. Therefore, they give rise to two ellipses of the axis
ratios
\begin{equation}
s_{1}=\sqrt{\frac{\mathcal{I}_{2}(\mathcal{I}_{2}-\mathcal{I}_{1})}{\mathcal{I}_{3}(\mathcal{I}_{3}-\mathcal{I}_{1})}},\;s_{3}=\sqrt{\frac{\mathcal{I}_{1}(\mathcal{I}_{3}-\mathcal{I}_{1})}{\mathcal{I}_{2}(\mathcal{I}_{3}-\mathcal{I}_{2})}}
\end{equation}
From the theory of rigid body dynamics, it is known that the dynamical
stability of the system decreases as these ratios deviate from a value
of one, and is greatest when a spinning top is symmetrical with respect
to the main axis of rotation; a case in which $s_{1}=1$ resp. $s_{3}=1$
applies. The reason for this is that the rotation around the main
axis can only be viewed as dynamically stable if the elliptical polhodia
adjacent to the point-shaped polhodia are not too eccentric, as otherwise
any originally adjacent pole could be pushed away from its initial
position after being affected by small perturbations; see here e.g.
\cite{arnol2013mathematical,grammel1920kraftefreie,klein2008theory,klein2010theory}
for overviews. 

The second of the above elliptic equations $(39)$ gives rise to the
so-called separatrix or separating polhodia, which is subject to the
relation $\ell^{2}=2\mathcal{H}\mathcal{I}_{2}$, thereby yielding
the result
\begin{equation}
\omega_{3}=\pm\sqrt{\frac{\mathcal{I}_{1}(\mathcal{I}_{2}-\mathcal{I}_{1})}{\mathcal{I}_{3}(\mathcal{I}_{2}-\mathcal{I}_{3})}}\omega_{1}.
\end{equation}
The separatrix consists of two plane curves which, as elliptical sections,
must be ellipses. These plane curves separate the epicycloidal and
pericycloidal regions from each other. 

In the symmetric case, in which the pairs of 'Bloch moments of inertia'
coincide, the form of the solution of the Bloch-type Euler equations
$(35)$ considerably simplifies. For instance, in case that $\mathcal{I}_{1}=\mathcal{I}_{2}$
applies, it can be concluded that $\omega_{3}=\omega_{30}=const.$
is satisfied, so that $(35)$ becomes

\begin{equation}
\ddot{\omega}_{j}=-\zeta^{2}\omega_{j}
\end{equation}
with $j=1,2$ and $\zeta=\omega_{30}\sqrt{\frac{\mathcal{I}_{3}-\mathcal{I}_{1}}{\mathcal{I}_{2}}\frac{\mathcal{I}_{2}-\mathcal{I}_{3}}{\mathcal{I}_{1}}}$.
The solution can thus be cast in the form

\begin{align}
\omega_{1}(t) & =\omega_{1(0)}\cos\zeta t,\\
\omega_{2}(t) & =\omega_{2(0)}\sin\zeta t,\nonumber \\
\omega_{3}= & \omega_{3(0)}=const.,\nonumber 
\end{align}
and it becomes clear that the quantities $\mathcal{H}$ and $\ell^{2}$
depicted in $(36)$ constitute first integrals of motion. The remaining
symmetric cases can then be treated along similar lines, and the solutions
have a form analogous to $(43)$.

The latter is of further relevance because the constants of motion
$\mathcal{H}$ and $\ell^{2}$ of the symmetric Euler top give rise
to associated conserved linear operators in the Bloch picture, from
now on to be referred to as operator-valued constants of motion. To
see this, the general $SU(d)$-case should be considered once again.
The integrability of this generic higher-dimensional case, which requires
a larger set of conserved quantities and happens to be altogether
more complex, will be proven in the following section of this work.
However, as can be quickly seen, integrability is related to the existence
of trace-free operators of the form
\begin{equation}
\mathcal{C}=\vec{\ell}\cdot\vec{\Lambda},\;\mathfrak{H}=H-h\textbf{1}_{d}=\frac{1}{2}\vec{\omega}\cdot\vec{\Lambda}
\end{equation}
which can be used to define the operator-valued constants of motion
$\mathfrak{O}_{k}:=\mathfrak{H}^{k}$ with $k\in\mathbb{N}$ as well
as $\mathfrak{E}=\frac{1}{2}\{\mathcal{C},\mathfrak{H}\}$, which,
as can readily be checked, meet the conditions $\dot{\mathfrak{H}}=\dot{\mathfrak{E}}=0$.
Similarly, each type of operator function defined with respect to
the operator-valued constants of motion depicted in $(44)$ again
represents a constant of motion, whose existence ensures the integrability
of $(33)$ in the standard Bloch sphere picture. Additionally, Casimir
operator invariants of the form 
\begin{equation}
\mathfrak{C}_{m}=\overset{N-1}{\underset{i_{1}i_{2},...,i_{m}=1}{\sum}}\kappa_{i_{1}i_{2}...i_{m}}\Lambda_{i_{1}}\Lambda_{i_{2}}...\Lambda_{i_{m}}
\end{equation}
can be defined, which constitute additional constants of motion. The
latter follows from the fact that $\kappa_{i_{1}i_{2}...i_{m}}$ is
a symmetric constant tensor, which makes it clear that $\dot{\mathfrak{C}}=0$
applies. 

To conclude this first section, however, let it be noted that the
decomposition of both the considered density operator and the Hamiltonian
depicted in $(3)$ need not be performed with respect to the generalized
Gell-Mann basis. Any other operator basis would have served the purpose
just about as well as long as it is guaranteed that the basis elements
form a Lie algebra. In particular, angular momentum generators or
any other basis of Casimir operators could have been used for the
calculation.

\section{Bloch Motions in Isolated Quantum Systems II: Integrability and Dynamical
Stability}

Having clarified the above, it may next be observed that the dynamics
of any closed quantum system obeying the von Neumann equation $(1)$
is characterized in the Bloch vector representation by equations,
whose solutions define pairs of canonical variables $\{q_{k},p_{k}\}$
in Bloch component phase space, by virtue of the fact that the definitions
$q_{k}\equiv\varrho_{k}$ and $p_{k}\equiv\dot{\varrho}_{k}$ are
used for the generalized positions and momenta. Since the motion of
the Bloch vector is constrained to the $(N-2)$-hypersphere $S^{N-2}\subset\mathbb{R}^{N-1}$,
the natural phase space for the Hamiltonian dynamics is the cotangent
bundle $\mathcal{P}\equiv T^{*}S^{N-2}$, equipped with a canonical
symplectic form $\Omega=\overset{N-1}{\underset{k=1}{\sum}}dq_{k}\wedge dp_{k}$
restricted to the constraint surface. Given that, similar to classical
Hamiltonian mechanics, evolution laws $(6)$ and $(12)$ give rise
to the Hamilton equations 
\begin{align}
\dot{q}_{k} & =\frac{\partial\mathcal{\mathscr{H}}}{\partial p_{k}}\\
\dot{p}_{k} & =-\frac{\partial\mathcal{\mathscr{H}}}{\partial q_{k}},
\end{align}
where $\mathscr{H}=T+V$ with $T(p_{1},p_{2},...,p_{N-1})=\frac{1}{2}\overset{N-1}{\underset{k=1}{\sum}}p^{2}_{k}$
and $V(q_{1},q_{2},...,q_{N-1})$ applies by definition. By defining
the vector $x\equiv(q_{1},q_{2},...,q_{N-1};p_{1},p_{2},...,p_{N-1})$,
the listed two dynamical equations can be combined into a single equation
of the form 

\begin{equation}
\dot{x}_{k}=f_{k}.
\end{equation}
Since the quantum system considered is a closed system, the associated
dynamical system is a conservative one. The Hamilton equations $(46)$
and $(47)$ thus provide the results $\dot{q}_{k}=p_{k}$ and $\dot{p}_{k}=F_{k}$,
where $F_{k}=-\frac{\partial\mathscr{H}}{\partial q_{k}}$ is a Newtonian-type
force term.

To fully understand the dynamics of the quantum system being studied,
the Hamiltonian $\mathscr{H}(x_{1},x_{2},...,x_{n})$ needs to be
specified. However, at least in the simple qubit case, this can readily
be achieved by taking the following into account: Observing that the
length of the Bloch vector $\mathcal{R}=\sqrt{\vec{\varrho}\vec{\varrho}}$
remains constant in the course of unitary evolution, it becomes clear
that relations $(1)$ resp. $(12)$ define a spherical harmonic oscillator
meeting the constraints $\ddot{\varrho}_{j}-\overset{3}{\underset{n=1}{\sum}}\mathcal{K}_{jn}\varrho_{n}=0,\;\overset{3}{\underset{n=1}{\sum}}\varrho^{2}_{n}-\mathcal{R}^{2}=0$.
In consequence, using the decomposition $\mathcal{K}_{ij}=\mathcal{K}_{i}\delta_{ij}$,
these two conditions can be combined and encoded into a Lagrangian
$\mathcal{L}(\varrho_{n},\dot{\varrho}_{n})$ of the form $\mathcal{L}=\frac{1}{2}\overset{3}{\underset{n,m=1}{\sum}}(\dot{\varrho}^{2}_{m}-\mathcal{K}_{m}\varrho^{2}_{m})+\frac{1}{2}\Lambda(\overset{3}{\underset{n=1}{\sum}}\varrho^{2}_{n}-\mathcal{R}^{2})$,
the latter of which can be rewritten in the form 
\begin{equation}
\mathcal{L}=\frac{1}{2}\overset{3}{\underset{m=1}{\sum}}(\dot{q}^{2}_{m}-\mathcal{K}_{m}q^{2}_{m})+\frac{1}{2}\Lambda(\overset{3}{\underset{n=1}{\sum}}q^{2}_{n}-\mathcal{R}^{2})
\end{equation}
in the specified canonical variables, where the occurring Lagrangian
multiplier can be determined to be $\Lambda=\frac{1}{\mathcal{R}^{2}}\overset{3}{\underset{l=1}{\sum}}(\dot{q}^{2}_{l}-\mathcal{K}_{l}q^{2}_{l})$.
This leads to the nonlinear Newtonian equations of motion
\begin{equation}
\ddot{q}_{m}=-\mathcal{K}_{m}q_{m}-q_{m}\overset{3}{\underset{l=1}{\sum}}(\frac{\dot{q}^{2}_{l}}{\mathcal{R}^{2}}-\mathcal{K}_{l}q^{2}_{l}),
\end{equation}
generally referred to as the Neumann model in the literature; see
here \cite{babelon2003introduction,bellon2005spectrum} for further
details. The Bloch-vector dynamics thus has a transparent geometric
interpretation: it is the geodesic flow on $S^{N-2}$ perturbed by
the potential $V(q)=\overset{3}{\underset{k=1}{\sum}}\mathcal{K}_{k}q^{2}_{k}$. 

Given this insight, the form of the resulting equations of motion
makes it clear that also a Bloch Hamiltonian of the form 
\begin{equation}
\mathscr{H}=\frac{1}{2}\overset{3}{\underset{\underset{m\neq n}{m,n=1}}{\sum}}J^{2}_{mn}+\frac{1}{2}\overset{3}{\underset{m=1}{\sum}}\mathcal{K}_{m}q^{2}_{m}
\end{equation}
can be set up in this context, where $J_{mn}=\overset{3}{\underset{i=1}{\sum}}\epsilon_{mni}\ell_{i}=q_{m}p_{n}-q_{n}p_{m}$
with $\ell_{i}=\overset{3}{\underset{j,k=1}{\sum}}\epsilon_{ijk}q_{j}p_{k}$
applies in the given qubit case. To check the integrability of the
resulting Hamiltonian equations of motion and to rewrite the equations
in the typical fashion known from the theory of integrable systems,
it may be assumed that $\mathcal{K}_{3}\leq\mathcal{K}_{2}\leq\mathcal{K}_{1}$
applies in this context. Defining then the objects $Q=(q_{k})$ and
$P=(p_{k})$, the Hamiltonian equations $(46)$ and $(47)$ become
\begin{equation}
P=\dot{Q}=\mathcal{J}Q,\quad\dot{P}=\mathcal{J}P+\mathcal{M}Q.
\end{equation}
Ultimately, using the fact that $\mathcal{M}\equiv\dot{\mathcal{J}}$
and thus $\dot{P}=\ddot{Q}=\mathcal{K}Q$ applies in this context,
it becomes clear that the form of said equation proves fully consistent
with that of equations $(7)$ and $(13)$, thereby showing that the
mentioned equations can readily be recast in a manifestly Hamiltonian
form. The Hamiltonian and the symplectic form have the symmetry $P\rightarrow P+\lambda Q$,
$Q\rightarrow Q$, which gives rise to a group under which a Hamiltonian
reduction can be performed. This can readily be checked by taking
into account that the matrix $J=(J_{mn})$ occurring in $(52)$ can
be rewritten in the form $J=Q^{\intercal}P-P^{\intercal}Q$. 

To prove the Liouville integrability of the resultant system of Hamiltonian
equations $(46)$ and $(47)$, a convenient method is to turn to the
Lax formulation of the theory. The very heart of this approach is
the Lax equation

\begin{equation}
\dot{L}=[M,L],
\end{equation}
which is defined with respect to the matrices

\begin{equation}
L(\lambda)=\mathcal{I}^{2}+\frac{1}{\lambda}J,\;M(\lambda)=\Omega+\lambda\mathcal{I},
\end{equation}
where $\lambda$ is a free constant spectral parameter and $\Omega_{ij}=\overset{3}{\underset{k=1}{\sum}}\epsilon_{ijk}\omega_{k}$
applies by definition. Using this ansatz, the Lax equation can be
recast in the form $\dot{L}(\lambda)-[M(\lambda),L(\lambda)]=[J,\mathcal{I}]+[\mathcal{I}^{2},\Omega]+\frac{1}{\lambda}(\dot{J}+[J,\Omega])=0$.
This only yields the equations of motion, however, since the first
two terms cancel each other out \cite{babelon2003introduction}. The
Liouville integrability of the Hamiltonian equations $(46)$ and $(47)$
then is a straightforward consequence of the existence of three independent
quantities 
\begin{equation}
F_{k}=q^{2}_{k}+\overset{3}{\underset{\underset{l\neq k}{l=1}}{\sum}}\frac{J^{2}_{kl}}{\mathcal{K}_{k}-\mathcal{K}_{l}},
\end{equation}
generally known as Uhlenbeck variables in the literature \cite{babelon2003introduction}.
In terms of these conserved quantities, the Hamiltonian of the theory
can be expressed as 
\begin{equation}
\mathscr{H}=\frac{1}{2}\overset{3}{\underset{j=1}{\sum}}\mathcal{K}_{j}F_{j}
\end{equation}
provided that the condition $\overset{3}{\underset{k=1}{\sum}}F_{k}=\mathcal{R}^{2}$
is met. For details, see e.g. \cite{babelon2003introduction}. The
structure of the two constants of motion $F_{2}=trL^{2}$ and $F_{3}=trL^{3}$
reveals that the Hamiltonian and the square of the Bloch angular momentum
defined in $(36)$ are actually constants of motion. The remaining
constants of motion, which ensure the integrability of the derived
Euler-Poinsot equations, can ultimately be determined by simple calculation
of the constant expressions $F_{k}=trL^{k}$. In further consequence,
this yields operator-valued constants of motion of the form $(44)$,
which ensure the integrability of $(33)$ and thus the sequential
integrability of $(1)$ and $(10)$, respectively. 

With that clarified, the question naturally arises as to whether these
results can be generalized to the higher-dimensional case of qudit
Hilbert spaces. To address this question, a rather natural approach
is to employ the Adler-Kostant-Symes scheme, which is typically employed
in the construction of solutions of the modified Yang-Baxter equation
\cite{babelon2003introduction}. This scheme, among other things,
exploits the fact that the Lax equation provides a flow on the coadjoint
orbits defined by the coadjoint action of the dual of the Lie algebra
$SU(d)$. The corresponding flow equation can be written in the form

\begin{equation}
\dot{L}=ad^{\ast}M\cdot L=[M,L],
\end{equation}
provided that $M=R(\nabla\mathscr{H})=\underset{a,b=1}{\overset{M}{\sum}}r^{ab}dh_{a}\Lambda_{b}$
with $dh_{a}=Tr(\nabla\mathscr{H}\Lambda_{a})$ applies, where $(r^{ab})$
is the so-called $r$-matrix. This shows that the matrix $M(\lambda)$
can be constructed from a suitable invariant Hamiltonian $\mathscr{H}$.
But this Hamiltonian, as may be noted, is not uniquely defined; there
are several Hamiltonians within the one and the same integrable family
that could be chosen in this context, all of which lead to different
types of dynamics.

To actually prove the Liouville integrability of the theory for arbitrary
$d$, it has to be ensured that a given Hamiltonian system exhibits
a set $\{F_{k}\}$ of $\frac{d(d-1)}{2}$ conserved quantities $F_{k}=\text{Tr}L^{k}$
with $F_{j}\equiv\mathscr{H}$ for $j\in\{1,2,...,\frac{d^{2}-1}{2}\}$
on an open, dense subset of $\mathcal{P}$. The elements of this set
have to meet the two criteria: $i)$ Involution: $\{F_{j},F_{k}\}=0$
for all $j,k\in\{1,2,...,\frac{N-1}{2}\}$ and $ii)$ Functional Independence:
$dF_{1}\wedge dF_{2}\wedge...\land dF_{\frac{d^{2}-1}{2}}\neq0$.
For $SU(d)$, the relevant phase space is the coadjoint orbit $\mathcal{O}(\ell_{0})=\{Ad^{*}_{g}(\ell_{0})\vert g\in SU(d)\}\subset\mathfrak{su}(d)^{*}$
with dimension $\text{dim\ensuremath{\mathcal{O}}(\ensuremath{\ell_{0}})}=d(d-1)$. 

That said, to make a suitable ansatz for the Lax pair $L(\lambda)$
and $M(\lambda)$, let it first be noted that the matrix $K=\overset{N-1}{\underset{k=1}{\sum}}\omega_{k}\Lambda_{k}$
constitutes a suitable choice for the matrix $M$ occurring in $(57)$,
i.e. $M\equiv K$. The corresponding Lax matrix then can be specified
using a Mishchenko-Fomenko shift argument ansatz of the form $L(\lambda)=\mathcal{C}+\lambda K$
with $\mathcal{C}=\overset{N-1}{\underset{k=1}{\sum}}\ell{}_{k}\Lambda_{k}$,
which shows that equation $(57)$ reduces to the form 
\begin{equation}
\frac{\partial\mathcal{C}}{\partial t}=[\mathcal{C},K].
\end{equation}
 This leads to equations $(32)$ and $(34)$ and, furthermore, gives
the Poisson bracket

\begin{equation}
\frac{\partial\ell_{k}}{\partial t}=\{\ell_{k},\mathscr{H}\}=\overset{N-1}{\underset{m,n=1}{\sum}}f_{kmn}\ell_{m}\omega_{n}
\end{equation}
with $\mathscr{H}=\frac{1}{2}TrLK=\frac{1}{2}\overset{N-1}{\underset{k=1}{\sum}}\ell_{k}\omega_{k}=\frac{1}{2}\overset{N-1}{\underset{k=1}{\sum}}\mathcal{I}_{k}\omega^{2}_{k}$.
Since $M$ does not depend on the spectral parameter, the corresponding
Sklyanin bracket simplifies to
\begin{equation}
\{L(\lambda)\overset{\otimes}{,}L(\mu)\}=\left[\frac{\mathcal{P}}{\lambda-\mu},L_{1}(\lambda)+L_{2}(\mu)\right]=\{\mathcal{C}\overset{\otimes}{,}\mathcal{C}\}
\end{equation}
thereby showing that the resultant expression is independent of the
concrete choice of spectral parameters $\lambda$ and $\mu$. As shown
in Appendix $A$, a direct computation using the Lie-Poisson structure
on $\mathfrak{su}(d)^{*}$ yields the Sklyanin bracket \cite{babelon2003introduction}
\begin{equation}
\{\mathcal{C}\overset{\otimes}{,}\mathcal{C}\}=[r,\mathcal{C}\otimes\textbf{1}_{d}+\textbf{1}_{d}\otimes\mathcal{C}]
\end{equation}
with the classical $r$-matrix $r=\frac{1}{2i}\Pi$, where $\Pi$
is the permutation operator on $\mathfrak{su}(d)\otimes\mathfrak{su}(d)$.
Since the $r$-matrix satisfies the classical Yang-Baxter equation
(see e.g. chapter $3$ of \cite{babelon2003introduction}), involution
of the spectral invariants
\begin{equation}
F_{k}(\lambda)=\text{Tr}L(\lambda)^{k},\{F_{k}(\lambda),F_{k}(\mu)\}=0
\end{equation}
follows by the Sklyanin theorem \cite{kulish1979quantum}. More specifically,
said involution follows from the Sklyanin bracket and, in further
consequence, from
\begin{equation}
\{\text{Tr}L(\lambda)^{k},\mathrm{\text{Tr}}L(\mu)^{l}\}=0
\end{equation}
 by standard arguments \cite{babelon2003introduction,kulish1979quantum,sklyanin1989separation}.
From the Lax pair, the conserved quantities emerge as coefficients
of the characteristic polynomial of $L(\lambda)$
\begin{equation}
\det(\mu\textbf{1}_{d}-L(\lambda))=\overset{d}{\underset{k=1}{\sum}}(-1)^{k}\sigma_{k}(\lambda)\mu^{d-k}=0,
\end{equation}
which gives rise to the spectral curve. The functions $\sigma_{k}(\lambda)$
occurring in this context are polynomials in $\lambda$ of the form
\begin{equation}
\sigma_{k}(\lambda)=\overset{d}{\underset{j=1}{\sum}}\lambda^{j}I_{k,j},
\end{equation}
where $I_{k,j}=\text{Tr}(\mathcal{C}^{k-j}K^{j})$ applies by definition.
These coefficients $I_{k,j}$ also occur in the definition of $F_{k}(\lambda)$,
i.e.
\begin{equation}
F_{k}(\lambda)=\text{Tr}L(\lambda)^{k}=\overset{d}{\underset{j=1}{\sum}}\left(\begin{array}{c}
k\\
j
\end{array}\right)\lambda^{j}I_{k,j},
\end{equation}
thereby making it clear that the latter thus constitute conserved
quantities in involution, where the latter, as shown in Appendix $A$
of this work, follows from the $r$-matrix structure. Functional independence
at the generic point follows from non-degeneracy of the spectral curve
$(64)$; see here \cite{adler1980p,hitchin1987stable}.

As should be noted in this context, the independence of the constructed
conserved quantities can be checked by counting dimensions on the
generic coadjoint orbit: The generic coadjoint orbit of $SU(d)$ has
dimension $d^{2}-d$ (since the stabilizer of a generic element is
a maximal torus subgroup of dimension $d-1$). According to the Arnold-Liouville
theorem \cite{arnol2013mathematical,babelon2003introduction,landau1982mechanics},
one thus requires $\frac{d{{}^2}-d}{2}$ independent conserved quantities
in involution. The spectral invariants $I_{k,j}$ with $k=2,...,d$
and $j=0,...,k$ yield $\sum^{d}_{k=2}(k-1)=\frac{d(d-1)}{2}$ candidates.
Among these, $d-1$ are trivially Casimirs, and others depend on trace
relations. The remaining independent variables are most readily counted
in the case of a regular semisimple element, where the spectral curve
defines a smooth algebraic curve and the number of independent leaves
is exactly $\frac{d^{2}-d}{2}$ - a result from the theory of algebraic
integrable systems \cite{adler1980p,hitchin1987stable}. This yields
exactly the number of independent integrals required for Liouville
integrability. 

Another approach to the subject would be to impose the zero-curvature
condition of an $\mathfrak{su}(d)$-valued connection
\begin{equation}
\partial_{t}L+\partial_{\lambda}M+[M,L]=0,
\end{equation}
which ensures the compatibility of two linear systems
\begin{equation}
\partial_{\lambda}\Psi=L\Psi,\;\partial_{t}\Psi=M\Psi,
\end{equation}
where $\Psi(\lambda,t)$ are the eigenfunctions of the Lax matrix
$L(\lambda)$. The monodromy data of the $\lambda$-system are then
conserved quantities of the time-dependent dynamics. For the von Neumann
equation on $\mathfrak{su}(d)^{*}$, the associated Lax pair in this
context can be chosen to be $L(\lambda)=\overset{N-1}{\underset{j=1}{\sum}}\frac{A^{(j)}}{\lambda-\lambda_{j}}$
and $M(\lambda)=\overset{N-1}{\underset{j=1}{\sum}}\frac{B^{(j)}}{\lambda-\lambda_{j}}=\overset{N}{\underset{j=1}{\sum}}\frac{1}{\lambda-\lambda_{j}}\underset{\underset{i\neq j}{i=1}}{\overset{N-1}{\sum}}\frac{A^{(j)}}{\lambda_{j}-\lambda_{i}}$.
This choice leads naturally to another approach to the construction
of generalized Uhlenbeck-type variables $F_{k}(\lambda)$, as may
be noted, consists in considering a Gaudin-type Hamiltonian $\mathscr{H}=\frac{1}{2}Tr(KL^{2})$,
which lies in the same integrable family \cite{babelon2003introduction}.
The Hamiltonian can then be written as a linear combination of the
residuals

\begin{equation}
F_{j}=\text{Res}_{\lambda=\lambda_{i}}\text{Tr}L(\lambda)^{2}=L_{jj}+\overset{N-.1}{\underset{\underset{k\neq j}{j=1}}{\sum}}\frac{L_{jk}L_{kj}}{\lambda_{j}-\mathcal{\lambda}_{k}},
\end{equation}
meaning that $\mathscr{H}=\frac{1}{2}\overset{N-1}{\underset{j=1}{\sum}}\lambda_{j}F_{j}$
applies, analogous to the case of the Neumann model. Another possibility
of constructing suitable spectral invariants is to use the method
of Hitchin \cite{hitchin1987stable}, which gives rise to conserved
quantities of the form
\begin{equation}
\mathscr{H}_{k}=\frac{1}{2i}\underset{\gamma}{\oint}\text{Tr}L(\lambda)^{k}d\lambda=\overset{M}{\underset{j=1}{\sum}}\text{Res}_{\lambda=\lambda_{j}}\text{Tr}L(\lambda)^{k}.
\end{equation}
The conserved quantities $I_{k,j}$ admit a natural interpretation
as 'Hitchin Hamiltonians' on the spectral curve $(64)$ associated
with the Lax matrix $L(\lambda)=\overset{N-1}{\underset{j=1}{\sum}}\frac{A^{(j)}}{\lambda-\lambda_{j}}$.
The spectral curve $\Sigma$ has genus $g=\frac{(d-1)(d-2)}{2}$,
and the objects $\mathscr{H}_{k}$ generate the complete integrable
system on the Jacobian $\text{Jac}(\Sigma)$ \cite{hitchin1987stable}.
To obtain the correct number of conserved quantities, a possible choice
for the objects $A^{(j)}$ would be $A^{(j)}=\alpha^{(j)}\mathcal{C}+\beta^{(j)}K$
instead of the standard choice of independent $A^{(j)}\in\mathfrak{su}(d)$.
The involution $\{\mathscr{H}_{k},\mathscr{H}_{l}\}=0$ then follows
from the geometry of $\text{Jac}(\Sigma)$, whereas the question of
Liouville integrability reduces to verifying the functional independence
of these invariants on the corresponding coadjoint orbit $\mathcal{O}$.
For a regular element $\omega\in\mathcal{O}$ with distinct eigenvalues
and a generic Bloch inertia operator $\mathcal{I}:=(\mathcal{I}_{jk})$,
the functions $I_{k,j}=\text{Tr}(\mathcal{C}^{k-j}K^{j})$ then form
a family whose functional independence can be established on an open
dense subset of the coadjoint orbit $\mathcal{O}$. That is to say,
one can select a subset of the Hitchin Hamiltonians $\{\mathscr{H}_{k}\}$
whose differentials are generically linearly independent, so that
the number of independent integrals matches half the orbit dimension.
Under these genericity assumptions $\text{rank(}\frac{\partial\mathscr{H}_{k}}{\partial\omega_{a}})=\frac{d(d-1)}{2}$can
be shown to hold by a direct evaluation of the differentials on the
tangent space $T_{x}\mathcal{O}$. In particular, for variations of
the form $\delta\mathcal{C}=[E_{ij},\mathcal{C}]$ that are formed
with respect to matrix units $\{E_{ij}\}$, the differentials produce
contributions proportional to $\lambda_{i}-\lambda_{j}$. These contributions
are multiplied by coefficients depending on the inverse $\mathcal{I}^{-1}$
of the Bloch inertia operator $\mathcal{I}$, where the scalars $\lambda_{i}$
are the eigenvalues of $\mathcal{C}$. For generic $\mathcal{I}^{-1}$,
these coefficients are non-degenerate, ensuring linear independence
and hence maximal rank. This establishes Liouville integrability of
the corresponding $SU(d)$ dynamical system.

As for the definition of time-dependent conserved quantities, the
method described in \cite{lewis1969exact} can be used to solve $(31)$
and $(58)$, which should yield time-dependent operator-valued constants
of motion of the type $\mathcal{C}(t)=\overset{N-1}{\underset{k,j=1}{\sum}}c_{kj}(t)I_{k,j}(t)=\overset{N-1}{\underset{k,j=1}{\sum}}c_{kj}\text{Tr}(\mathcal{C}(t)^{k-j}K(t)^{j})$.
In exactly the same way as above, all other required operator-valued
constants of motion can then be constructed using a series expansion
in generalized Gell-Mann bases and repeating all the steps outlined
above. The spectral invariants $F_{k}(\lambda,t)=\mathrm{Tr}L(\lambda,t)^{k}$
are then isomonodromic invariants resulting from $(66)$ - not conserved
quantities in the classical sense, but constant along the isomonodromic
flow.

This, in combination with the above, shows that the Bloch dynamics
of closed quantum systems (including the time-dependent case) are
Liouville-integrable for any (including time-dependent) Hamiltonians
$H(t)$, since the latter can be reduced to the autonomous Euler-Poinsot
system $(32)$, whose integrability can be proven via the Lax pair
$(L(\lambda),K)$. Yet, as may be noted, the reconstruction of $\vec{\rho}(t)$
from $\vec{\omega}(t)$ via $(7)$ requires additional integration.
Naturally, the solvability in closed form thus depends on the specific
form of $H(t)$. The integrability of $(32)$ resp. $(58)$ and $(59)$
is therefore a necessary but not sufficient condition for the integrability
of the entire system in the time-dependent case. It may very well
be possible that no closed-form solution exists in specific cases.
In the principal axis system, however, the problem is completely Liouville-integrable
for any $H(t)$. In this case, the Dyson series $(93)$ provides a
suitable approach. Concretely, using $U(t)=\mathcal{T}\exp\{-\overset{t}{\underset{0}{\int}}K(\tau)d\tau\}=\mathcal{T}\exp\{-i\overset{t}{\underset{0}{\int}}H(\tau)d\tau\}$
in combination with $\mathcal{C}(t)=U(t)\mathcal{C}(0)U^{\dagger}(t)$,
$H(t)=U(t)H(0)U^{\dagger}(t)$ and $\rho(t)=U(t)\rho(0)U^{\dagger}(t)$,
it can be shown that 
\begin{equation}
\ell_{k}(t)=\overset{N-1}{\underset{j=1}{\sum}}R_{kj}(t)\ell_{j}(0),\;\rho_{k}(t)=\overset{N-1}{\underset{j=1}{\sum}}R_{kj}(t)\rho_{j}(0)
\end{equation}
with 
\begin{equation}
R_{kj}(t)=\frac{1}{2}\text{Tr}\left(\Lambda_{j}U(t)\Lambda_{k}U^{\dagger}(t)\right).
\end{equation}
By the Arnold-Liouville theorem it is known that action angle coordinates
and a linear flow exist on the solution Tori. The latter, however,
can be used to derive a closed-form exact solution for the system
$(71)$. The explicit representation of this solution can be given
in terms of the Baker-Akhiezer function \cite{babelon2003introduction,belokolos1994algebro,krichever1977methods}.
To obtain said function, the first step is to calculate the eigenvalues
$\mu(\lambda)$ and eigenfunctions $\psi(\lambda,t)$ of the Lax matrix
$L(\lambda)$, that is, to solve $\det(L(\lambda)-\mu\textbf{1})=0$
and $L(\lambda)\psi(\lambda,t)=\mu\psi(\lambda,t)$. The second step
consists of making sure that the eigenfunctions $\psi(\lambda,t)$
obey the first order differential relation $\partial_{t}\psi(\lambda,t)=M\psi(\lambda,t)$
and to identify the poles $\mathcal{D}(0)$ of $\psi(\lambda,t)$.
The third step is to use the asymptotic expansion of $M(\lambda)$
to show that there exists a suitable meromorphic differential $d\Omega$
such that $\chi_{j}(\lambda):=\overset{P}{\underset{P_{0}}{\int}}\omega_{j}$
with respect to a given point $P(\lambda)=(\mu(\lambda),\lambda)$
on the spectral curve $\Sigma$. By identifying then $\mathcal{O}\rightarrow\text{Jac}(\Sigma)$
and using the fact that $\frac{d}{dt}(\det(L(\lambda)-\mu\textbf{1})=0$
is fulfilled in this context \cite{babelon2003introduction}, it is
known that there exists a Baker-Akhiezer function given by the Its-Matveev
formula \cite{its1975schrodinger,krichever1977methods}
\begin{equation}
\psi_{j}(P,t)=\frac{\Theta(\mathcal{\vec{A}}(P)+\vec{\mathcal{A}}(\mathcal{D}(t))+\vec{K}+\vec{\Delta}_{j}\vert\tau)}{\Theta(\mathcal{\vec{A}}(\mathcal{D}(0))+\vec{K}_{j}\vert\tau)}\exp t\chi_{j},
\end{equation}
where $\Theta(\vec{z}\vert\tau)=\underset{\vec{n}\in\mathbb{C}^{g}}{\sum}\exp(i\pi\vec{n}^{\intercal}\tau\vec{n}+i\pi\vec{n}^{\intercal}\vec{z})$
is the Riemann theta function with half-period ratio matrix $\tau\in\mathbb{C}^{g\times g}$
with $\tau_{jk}=\underset{b_{j}}{\oint}\omega_{k}$ associated with
the spectral curve $\Sigma$, $\mathcal{\vec{A}}(P)=(\overset{P}{\underset{P_{0}}{\int}}\omega_{1},...,\overset{P}{\underset{P_{0}}{\int}}\omega_{g})$
and $\vec{K}_{j}$ is vector-valued initial data. Using then that
the exact linearization of the flow on the Jacobian via the Abel-Jacobi
map applies exactly in the given case, i.e. $\vec{\mathcal{A}}(\mathcal{D}(t))=\vec{\mathcal{A}}(\mathcal{D}(0))+\vec{\Omega}t$,
where $\vec{\Omega}\in\mathbb{C}^{g}$ is a constant frequency vector,
the equivalent relation
\begin{equation}
\partial_{t}\ln\psi_{j}(P,t)=\omega_{j}+\partial_{t}\log\frac{\Theta(\vec{u}(t)+\vec{c}_{j}\vert\tau)}{\Theta(\vec{u}(t)\vert\tau)}
\end{equation}
can be derived, where $\vec{c}_{j}(P,P_{0}):=\mathcal{\vec{A}}(P_{j})-\mathcal{\vec{A}}(P_{0})+\vec{\Delta}_{j}$
and $\vec{u}(t)=\vec{u}_{0}+\vec{\Omega}t$ with $\vec{u}_{0}=\mathcal{\vec{A}}(\mathcal{D}(0))+\vec{K}$
applies by definition. 

Given the above, it becomes clear the Lax operator can be diagonalized
via $L=\Psi\Lambda\Psi^{-1}$ with $\Lambda=\text{diag}(\mu_{1},\mu_{2},...,\mu_{d})$
and $\Psi(\lambda,t)=(\psi_{1}(\lambda,t),\psi_{2}(\lambda,t),...,\psi_{d}(\lambda,t))$,
thereby yielding $\mathcal{C}=\Psi_{0}\Lambda_{0}\Psi^{-1}_{0}$ with
$\Psi_{0}(t):=\Psi(\lambda,t)\vert_{\lambda=0}$ and $\Lambda_{0}:=\Lambda(\lambda)\vert_{\lambda=0}.$
This gives the result 
\begin{equation}
\ell_{k}(t)=\text{tr}(\mathcal{C}(t)\Lambda_{k})=\text{tr}(\Psi_{0}(t)\Lambda_{0}\Psi^{-1}_{0}(t)\Lambda_{k}).
\end{equation}
By introducing Sklyanin separation variables $(\lambda_{j},\mu_{j})$
with $j\in\{1,2,...,g\}$, which are defined as the zeroes of a specific
entry of the Lax matrix, this relation can be slightly rewritten.
More specifically, for the Bloch system with Lax matrix $L(\lambda)=\mathcal{C}+\lambda K$,
the standard method is to consider some suitable off-diagonal entry
$L_{ij}(\lambda)$ with $i\neq j$ (typically the $(1,2)$-entry)
and to define a function of the form $Q(\lambda)=\mathcal{C}{}_{ij}+\lambda K_{ij}$
for fixed $i,j\in\{1,2,...,g\}$. The separation variables are then
the zeroes $\lambda_{k}$ of $Q(\lambda)$, i.e. $Q(\lambda)\vert_{\lambda=\lambda_{k}}=0$
and thus $\lambda_{k}=-\frac{\mathcal{C}{}_{ij}}{K_{ij}}$. The remaining
variables read $\mu_{k}=\mathcal{C}{}_{ii}+\lambda_{k}K_{ii}$. Combining
these two variables results in pairs $(\lambda_{k},\mu_{k})$ that,
by definition, lie on the spectral curve $\Sigma$, i.e. $\det(L(\lambda_{k})-\mu_{k}\textbf{1})=0$.
As it turns out, these separation variables are canonically conjugated
in the sense that $\{\lambda_{j},\mu_{k}\}=\delta_{jk}$, $\{\lambda_{j},\lambda_{k}\}=\{\mu_{j},\mu_{k}\}=0$
and are subject to the equations of motion $\partial_{t}\lambda_{j}=\{\lambda_{j},H\}=\frac{\partial H}{\partial\mu_{j}}\vert_{\Sigma}$
and $\partial_{t}\mu_{j}=\{\mu_{j},H\}=\frac{\partial H}{\partial\lambda_{j}}\vert_{\Sigma}$.
Using these variables, one can define objects $u_{k}(t)=\overset{g}{\underset{j=1}{\sum}}\overset{(\lambda_{j}(t),\mu_{j}(t))}{\underset{P_{0}}{\int}}\omega_{k}$
with the property that $\partial_{t}u_{k}=\partial_{t}\overset{g}{\underset{j=1}{\sum}}\overset{(\lambda_{j}(t),\mu_{j}(t))}{\underset{P_{0}}{\int}}\omega_{k}=:\Omega_{k}=\text{const}.$,
thus making it clear that $u_{k}(t)=u_{k}(0)+\Omega_{k}t$. For given
$u_{k}(t)\in\text{Jac}(\Sigma)$, one can thus, conversely, formulate
a Jacobi inversion problem that aims to find suitable pairs $(\lambda_{j}(t),\mu_{j}(t))$
such that $u_{k}(t)$ is linear in $t\in\mathbb{R}$, thereby ensuring
the linearization of the Abel-Jacobi flow on $\text{Jac}(\Sigma)$.
The solution to the Jacobi inversion problem is given by the zeros
of the function $F(P):=\Theta\left(\vec{\mathcal{A}}(P)-\vec{u}(t)-\vec{K}|\tau\right)$.
By exploiting the degenerate form of Fay's trisecant identity \cite{fay2006theta},
logarithmic derivatives of theta-function ratios can be expressed
as meromorphic functions with prescribed divisor. In particular, the
observables $\ell_{k}(t)$ being rational functions of the Lax matrix,
inherit this structure. They can therefore be represented in the form
\begin{equation}
\ell_{k}(t)=\bar{\ell}_{k}+\partial_{u_{k}}\ln\frac{\Theta(\vec{u}(t)+\vec{c}_{k}\vert\tau)}{\Theta(\vec{u}(t)\vert\tau)},
\end{equation}
where it has been used that $\partial_{t}\log\frac{\Theta(\vec{u}(t)+\vec{c}_{k}\vert\tau)}{\Theta(\vec{u}(t)\vert\tau)}=\Omega_{k}\partial_{u_{k}}\ln\frac{\Theta(\vec{u}(t)+\vec{c}_{k}\vert\tau)}{\Theta(\vec{u}(t)\vert\tau)}$
applies in the given context. A proof of this relation is given in
Appendix $A$ of the work. Here, as may be noted, the quantity $\bar{\ell}_{k}=\frac{1}{T}\overset{T}{\underset{0}{\int}}\ell_{k}(t)dt$
constitutes the time-averaged value of the $k$-th Bloch component.
Since $\rho_{k}(t)$ can be reconstructed from $\mathcal{C}(t)$,
as may be noted, it inherits the same theta-functional structure.
One should obtain an expression for $\rho_{k}(t)$ analogous to $(76)$,
thereby giving the complete explicit exact closed-form solution to
the Euler-Poinsot-Bloch system for general $SU(d)$. The main details
of the derivation of this solution, as may be noted, will be given
in a future work on the subject.

That being said, let the following be noted: It is straightforward
to introduce a stability criterion for quantum dynamics within the
treated Bloch vector formalism simply by considering a Bloch analogue
of the Routh-Hurwitz criterion of classical Newtonian dynamics. The
latter is a necessary and sufficient criterion for the linear stability
of dynamical systems, which states that a dynamical system is stable
if and only if all eigenvalues of the Jacobi matrix occurring in the
equations of motion have a strictly negative real part, thereby requiring
the Jacobi matrix to be a Hurwitz stability matrix. To lift this criterion
into the given physical context, it may therefore be required that
all eigenvalues $\mathcal{J}_{k}$ of the Jacobi-type matrix $\mathcal{J}\equiv(\mathcal{J}_{kl})$
have strictly negative real part, i.e. 
\begin{equation}
\Re(\mathcal{J}_{k})<0.
\end{equation}
The problem with this criterion, however, is that it is automatically
fulfilled in the given context due to the fact that the matrix $\text{\ensuremath{\mathcal{J}}}$is
skew-symmetric. Therefore, applying $(77)$ to unitary dynamics does
not yield any additional insights, since the latter, by definition,
are linearly stable. The latter, however, is in stark contrast to
the non-unitary dynamics of open quantum systems, which, in the Markovian
case, are described by the Lindblad equation. Applying condition $(77)$
to such non-unitary dynamics could very well yield new insights and
warrants future investigation.

To show nonlinear stability, it generally proves more convenient to
additionally use the Energy-Casimir method discussed in \cite{aeyels1992stabilization,bloch1990stabilization},
which applies to Hamiltonian systems with Lie-Poisson structure and
thus also to the aforementioned Euler-Poinsot system. This Energy-Casimir
method makes use of a function of the form 
\begin{equation}
\mathfrak{Y}=\mathcal{H}+\overset{M}{\underset{k=1}{\sum}}\mu_{k}\mathscr{C}_{k},
\end{equation}
where $\mathscr{C}_{k}=tr\mathfrak{H}^{k}$ are Casimir functions,
i.e. invariants of the operator $\mathfrak{O}_{k}:=\mathfrak{H}^{k}$
and $\mu_{k}$ are Lagrange multipliers. More specifically, the method
checks the positive definiteness of the second variation (sometimes
referred to as Hessian form) related to $\mathfrak{Y}$, restricted
to the coadjoint orbit $\mathcal{O}(\omega_{0})=\{Ad^{*}_{g}(\omega_{0})\vert g\in G\}\subset\mathfrak{su}(d)^{*}$
through a generic point $\omega_{0}\subset\mathfrak{su}(d)^{*}$,
where the latter is defined with respect to the Lie group $G$ associated
with the examined Lie algebra $\mathfrak{g}\equiv SU(d)$. The mentioned
coadjoint orbit through a generic element has dimension $d(d-1)$,
so that the phase space has dimension $\frac{d(d-1)}{2}$ in the sense
of Liouville. By restriction to this coadjoint orbit, a system configuration
can be determined for which a specific critical point $\omega_{0}$
constitutes a strict local extremum, that is, a minimum of $\mathfrak{Y}$
on the tangent space $T_{\omega_{0}}\mathcal{O}\subset\mathfrak{g}^{*}$
of a given symplectic leaf $\mathcal{O}(\omega_{0})$. As a basis
for this, the observation is used that whenever $\mathfrak{Y}$ remains
constant on the mentioned coadjoint surfaces and suitable Lagrange
multipliers $\mu_{k}$ are selected to set up a positive definite
Hessian form $\delta^{2}\mathfrak{Y}=h_{jk}\delta\omega_{j}\delta\omega_{k}$
with 

\begin{equation}
h_{jk}=\frac{\partial^{2}\mathfrak{Y}}{\partial\omega_{j}\partial\omega_{k}}\vert_{\omega_{0}},
\end{equation}
the local minimum should give rise to a region in Bloch phase space
that cannot be left; so that the system must be Lyapunov stable. Such
a system configuration can be specified by demanding the validity
of $\delta\mathfrak{Y}\equiv\frac{\partial^{2}\mathfrak{Y}}{\partial\omega_{i}}\vert_{\omega_{0}}\delta\omega_{i}=0$,
which is tantamount to requiring
\begin{equation}
\nabla_{\omega_{i}}\mathfrak{Y}\vert_{\omega_{0}}=\frac{\partial\mathfrak{Y}}{\partial\omega_{i}}\vert_{\omega_{0}}=0,
\end{equation}
where $\omega_{0}$ is a critical point of the system. Though technically
demanding in general, this approach offers the possibility of rigorously
identifying system configurations that are stable in a nonlinear sense.
To show this, the $SU(3)$-case is discussed as a concrete example
in section four of this paper. On the other hand, a more convincing
and conclusive derivation for the entire $SU(d)$-case will only be
provided elsewhere.

\section{Bloch Motions in Composite Quantum Systems}

The arguments laid out in the previous section, which were only invoked
in respect to isolated quantum systems, shall now be extended to the
more complicated case of composite quantum systems. To this end, it
makes sense to first consider the simplest possible case of a bipartite
quantum system defined with respect to an associated bipartite Hilbert
space $\mathcal{H}=\mathcal{H}_{1}\otimes\mathcal{H}_{2}$. In this
bipartite case, any density operator and any Hamiltonian can be decomposed
according to the rule

\begin{align}
\rho & =\rho_{00}\textbf{1}_{d}\otimes\textbf{1}_{d}+\overset{N-1}{\underset{j=0}{\sum}}\rho_{j0}\lambda_{j}\otimes\textbf{1}_{d}+\overset{N-1}{\underset{k=0}{\sum}}\rho_{0k}\textbf{1}_{d}\otimes\lambda_{k}+\overset{N-1}{\underset{j,k=0}{\sum}}\rho_{jk}\lambda_{j}\otimes\lambda_{k},\\
H & =h_{00}\textbf{1}_{d}\otimes\textbf{1}_{d}+\overset{N-1}{\underset{j=0}{\sum}}h_{j0}\lambda_{j}\otimes\textbf{1}_{d}+\overset{N-1}{\underset{k=0}{\sum}}h_{0k}\textbf{1}_{d}\otimes\lambda_{k}+\overset{N-1}{\underset{j,k=0}{\sum}}h_{jk}\lambda_{j}\otimes\lambda_{k},\nonumber 
\end{align}
where $\rho_{00}\equiv\rho_{0}\equiv\frac{1}{N}=const.$ and $h_{00}\equiv h_{0}$
applies by definition. As may be noted, the introduced matrix basis
is again a generalized Gell-Mann basis consisting of elements that
meet the conditions

\begin{align}
\{\lambda_{i},\lambda_{j}\} & =\frac{4}{d}\delta_{ij}\textbf{1}_{d}+2\overset{N-1}{\underset{k=1}{\sum}}\mathrm{g}_{ijk}\lambda_{k},\\{}
[\lambda_{i},\lambda_{j}] & =2i\overset{N-1}{\underset{k=1}{\sum}}\mathrm{f}_{ijk}\lambda_{k},\nonumber 
\end{align}
and the decomposition relations thus obtained are manifestly identical
to decomposition relations $(3)$. Combining the commutator relation
occurring in $(82)$ with decomposition relation $(81)$, it thus
becomes clear that also in the given case evolution equations for
the corresponding Bloch components can be derived. Using the operator
identity $[A\otimes B,C\otimes D]=\frac{1}{2}\left([A,B]\otimes\{C,D\}+\{A,B\}\otimes[C,D]\right)$
with $A,B,C,D\in\mathcal{B}(\mathcal{H})$ for deriving these equations,
the von Neumann equation $(1)$ gives the following four equations
for the Bloch components 
\begin{align}
\dot{\rho}_{00} & =0,\;\dot{\rho}_{i0}=\overset{N-1}{\underset{l=1}{\sum}}\mathcal{L}^{(1)}_{il}\rho_{l0}+\mathcal{M}^{(1)}_{i},\;\dot{\rho}_{0k}=\overset{N-1}{\underset{l=1}{\sum}}\mathcal{L}^{(2)}_{lk}\rho_{0l}+\mathcal{M}^{(2)}_{k},\\
\dot{\rho}_{jk} & =\overset{N-1}{\underset{l,m=1}{\sum}}\mathcal{N}_{jklm}\rho_{lm}+\mathcal{O}_{jk},\nonumber 
\end{align}
where details of the calculation are given in Appendix $B$ of this
work. Consequently, the Bloch equations of motion resulting from calculating
the time derivative of $(83)$ take the form
\begin{align}
\ddot{\rho}_{00} & =0,\;\ddot{\rho}_{i0}=\overset{N-1}{\underset{l=1}{\sum}}\mathcal{Q}^{(1)}_{il}\rho_{l0}+\mathcal{R}^{(1)}_{i},\;\ddot{\rho}_{0k}=\overset{N-1}{\underset{l=1}{\sum}}\mathcal{Q}^{(2)}_{lk}\rho_{0l}+\mathcal{R}^{(2)}_{k},\\
\ddot{\rho}_{jk} & =\overset{N-1}{\underset{l,m=1}{\sum}}\mathcal{S}_{jklm}\rho_{lm}+\mathcal{T}_{jk},\nonumber 
\end{align}
with details being given again in Appendix $B$ of the paper. Based
on the fact that decomposition relations $(3)$ and $(81)$ are identical,
it thus becomes clear that the resulting complex dynamics are intimately
related with the more compact and transparent overall dynamics of
the total system encoded in relations $(6)$ and $(12)$. Against
this background, it stands to reason to also consider the basis decomposition
\begin{equation}
\mathcal{C}=\overset{N-1}{\underset{j=0}{\sum}}\ell_{j0}\lambda_{j}\otimes\textbf{1}_{d}+\overset{N-1}{\underset{k=0}{\sum}}\ell_{0k}\textbf{1}_{d}\otimes\lambda_{k}+\overset{N-1}{\underset{j,k=0}{\sum}}\ell_{jk}\lambda_{j}\otimes\lambda_{k}
\end{equation}
of the linear operator defined in $(20)$. Combining this ansatz with
that for the Hamiltonian depicted in $(81)$ and inserting it into
equation $(31)$ yields the bipartite Bloch equations

\begin{align}
\dot{\ell}_{i0} & =\overset{N-1}{\underset{l=1}{\sum}}\mathcal{L}^{(1)}_{il}\ell_{l0}+\mathcal{U}^{(1)}_{i},\;\dot{\ell}_{0k}=\overset{N-1}{\underset{l=1}{\sum}}\mathcal{L}^{(2)}_{lk}\ell_{0l}+\mathcal{U}^{(2)}_{k},\\
\dot{\ell}_{jk} & =\overset{N-1}{\underset{l,m=1}{\sum}}\mathcal{N}_{jklm}\ell_{lm}+\mathcal{V}_{jk},\nonumber 
\end{align}
where the additional definitions are laid out in Appendix $B$ of
this work. These relations, in turn, are in one-to-one correspondence
with the Bloch vector equations $(32)$ of the overall system, which,
however, implies that also in the given case there should be operator-valued
constants of motion of the form $(44)$, whose existence ensures the
integrability of $(86)$. Furthermore, the integrability of $(83)$
is guaranteed by the results of the previous section. 

Remarkably, it can be observed that the Bloch equations (and thus
the dynamics of the individual quantum systems) fully decouple in
the event that $h_{jk}=0$ applies and an ansatz of the form
\begin{equation}
\rho_{jk}=\alpha^{(1)}_{j}\alpha^{(2)}_{k}
\end{equation}
is made for $\rho_{jk}$ in $(81)$; that is, provided that $\rho$
is a product state. As clarified in the relevant literature on the
entanglement detection problem \cite{de2007separability,hassan2008separability},
the stricter condition $\rho_{jk}=\overset{\mu}{\underset{\nu=1}{\sum}}\alpha^{(1)}_{j\nu}\alpha^{(2)}_{k\nu}$
constitutes a sufficient condition for a state to be separable, but
not a necessary one. Yet, as is well known, there are no necessary
and sufficient criteria for entanglement detection; with the exception
of the Peres-Horodecki criterion \cite{horodecki1997separability,peres1996separability},
which is such a criterion, but only in low dimensions (that is, in
the given case, for $d=2$). Still, it may be conjectured that separability
is a necessary prerequisite for the decoupling the dynamics of quantum
systems, though it is unclear whether $\rho$ actually has to be a
product state. Using the mentioned Peres-Horodecki criterion
\begin{equation}
\rho^{T_{j}}\geq0
\end{equation}
where $T_{j}$ (with either $j=1$ or $j=2$) denotes partial transposition
with respect to either the first or the second subsystem, it can also
be concluded from the low-dimensional case $N=4$ that the Bloch dynamics
for separable states are simpler than for entangled states due to
the imposed constraints. Showing this in detail, however, is left
to future work.

That said, with the clarification of the above points, all the groundwork
has been laid for handling the multipartite case. As is to be expected,
however, the situation in the case mentioned is more complicated,
i.e., when considering a composite Hilbert space $\mathcal{H}\equiv\otimes^{n}_{i=1}\mathcal{H}_{i}$
consisting of a finite number $n$ of $d$-level systems. This can
readily be seen as follows: Given an operator set $\{\lambda{}^{(0)},\lambda{}^{(j_{1})}_{k_{j_{1}}},...,\lambda{}^{(j_{1}j_{2}...j_{n})}_{k_{j_{1}},k_{j_{2}},...k_{j_{n}}}\}$
defined with respect to expressions of the form $\lambda{}^{(0)}:=\mathbf{1}_{d}\otimes\mathbf{1}_{d}\otimes...\otimes\mathbf{1}_{d}$,
$\lambda{}^{(j_{1})}_{k_{j_{1}}}:=\mathbf{1}_{d}\otimes\mathbf{1}_{d}\otimes...\otimes\lambda_{k_{j_{1}}}\otimes\mathbf{1}_{d}\otimes...\otimes\mathbf{1}_{d}$,
$\mathfrak{\lambda}{}^{(j_{1}j_{2})}_{k_{j_{1}}k_{j_{2}}}:=\mathbf{1}_{d}\otimes\mathbf{1}_{d}\otimes...\otimes\lambda_{k_{j_{1}}}\otimes\mathbf{1}_{d}\otimes...\otimes\mathbf{1}_{d}\otimes\lambda_{k_{j_{2}}}\otimes\mathbf{1}_{d}\otimes...\otimes\mathbf{1}_{d}$,...,
$\lambda{}^{(j_{1}j_{2}...j_{n})}_{k_{j_{1}},k_{j_{2}},...k_{j_{n}}}:=\lambda_{k_{j_{1}}}\otimes\lambda_{k_{j_{2}}}\otimes...\otimes\text{\ensuremath{\lambda}}_{k_{j_{n}}}$,
the density operator and the Hamiltonian can be expanded according
to the rule
\begin{align}
\rho & =\frac{1}{N}\lambda{}^{(0)}+\overset{}{\underset{\{j_{1}\}}{\sum}}\overset{N-1}{\underset{k_{j_{1}}=1}{\sum}}\rho_{k_{j_{1}}}\lambda{}^{(j_{1})}_{k_{j_{1}}}+\overset{}{\underset{\{j_{1},j_{2}\}}{\sum}}\overset{N-1}{\underset{k_{j_{1}},k_{j_{2}}=1}{\sum}}\rho_{k_{j_{1}}k_{j2}}\lambda{}^{(j_{1}j_{2})}_{k_{j_{1}}k_{j_{2}}}+...\\
 & ...+\overset{}{\underset{\{j_{1},j_{2},...,j_{2n}\}}{\sum}}\overset{N-1}{\underset{k_{j_{1}},k_{j_{2}},...k_{j_{m}}=1}{\sum}}\rho_{k_{j_{1}}k_{j_{2}}...k_{j_{m}}}\lambda{}^{(j_{1}j_{2}...j_{m})}_{k_{j_{1}},k_{j_{2}},...k_{j_{m}}}+...\nonumber \\
 & ...+\overset{N-1}{\underset{k_{j_{1}},k_{j_{2}},...k_{j_{n}}=1}{\sum}}\rho_{k_{j_{1}}k_{j_{2}}...k_{j_{n}}}\lambda{}^{(j_{1}j_{2}...j_{n})}_{k_{j_{1}},k_{j_{2}},...k_{j_{n}}}\nonumber 
\end{align}
and 
\begin{align}
H & =h_{0}\lambda{}^{(0)}+\overset{}{\underset{\{j_{1}\}}{\sum}}\overset{N-1}{\underset{k_{j_{1}}=1}{\sum}}h_{k_{j_{1}}}\lambda{}^{(j_{1})}_{k_{j_{1}}}+\overset{}{\underset{\{j_{1},j_{2}\}}{\sum}}\overset{N-1}{\underset{k_{j_{1}},k_{j_{2}}=1}{\sum}}h_{k_{j_{1}}k_{j2}}\lambda{}^{(j_{1}j_{2})}_{k_{j_{1}}k_{j_{2}}}+...\\
 & ...+\overset{}{\underset{\{j_{1},j_{2},...,j_{2n}\}}{\sum}}\overset{N-1}{\underset{k_{j_{1}},k_{j_{2}},...k_{j_{m}}=1}{\sum}}h_{k_{j_{1}}k_{j_{2}}...k_{j_{m}}}\lambda{}^{(j_{1}j_{2}...j_{m})}_{k_{j_{1}},k_{j_{2}},...k_{j_{m}}}+...\nonumber \\
 & ...+\overset{N-1}{\underset{k_{j_{1}},k_{j_{2}},...k_{j_{n}}=1}{\sum}}h_{k_{j_{1}}k_{j_{2}}...k_{j_{n}}}\lambda{}^{(j_{1}j_{2}...j_{n})}_{k_{j_{1}},k_{j_{2}},...k_{j_{n}}},\nonumber 
\end{align}
where $N=d^{n}$, $1\leq m\leq n$ holds and the abbreviations $\rho_{k_{j_{1}}}:=\frac{d}{2}Tr(\rho\lambda{}^{(j_{1})}_{k_{j_{1}}})$,
$\rho_{k_{j_{1}}k_{j_{2}}}:=\frac{d^{2}}{4}Tr(\rho\lambda{}^{(j_{1}j_{2})}_{k_{j_{1}}k_{j_{2}}})$,...,$\rho_{k_{j_{1}}k_{j_{2}}...k_{j_{n}}}:=\frac{N}{2^{n}}Tr(\rho\lambda{}^{(j_{1}j_{2}...j_{n})}_{k_{j_{1}},k_{j_{2}},...k_{j_{n}}})$
are used. Here, $\{j_{1},j_{2},...,j_{m}\}$ with $2\leq m\leq n$
is a subset of the set $\mathcal{N}=\{1,2,...,n\}$ and can be chosen
in $\left(\begin{array}{c}
n\\
m
\end{array}\right)$ ways, contributing $\left(\begin{array}{c}
n\\
m
\end{array}\right)$ terms in the sum $\Sigma_{\{j_{1},j_{2},...,j_{m}\}}$ in equations
$(89)$ and $(90)$, each containing a tensor of order $m$. The total
number of terms in the Bloch representation of $\rho$ and $H$ is
thus $2^{n}$. 

Using the above relations, evolution equations for the Bloch components
can be derived from $(1)$ and $(10)$ that generalize the bipartite
equations $(83)$ and $(84)$ to the multipartite case. The first
order equations take the form
\begin{equation}
\dot{\rho}_{k_{j_{1}}k_{j_{2}}...k_{j_{n}}}=\overset{N-1}{\underset{k_{j_{1}}...k_{j_{n}},l_{m_{1}}...l_{m_{n}}=1}{\sum}}\mathcal{N}_{k_{j_{1}}k_{j_{2}}...k_{j_{n}}l_{m_{1}}l_{m_{2}}...l_{m_{n}}}\rho_{l_{m_{1}}l_{m_{2}}...l_{m_{n}}}+\mathcal{O}_{k_{j_{1}}k_{j_{2}}...k_{j_{n}}}
\end{equation}
for all possible $n$, where $\mathcal{O}_{k_{j_{1}}k_{j_{2}}...k_{j_{n}}}$
is a lengthy, rather unilluminating expression that depends in a fairly
complicated way on the structure coefficients $\mathrm{f}_{ijk}$
and $\mathrm{g}_{ijk}$, the Bloch components of the Hamiltonian $h_{0}$,
$h_{k_{j_{1}}}$, ...., $h_{k_{j_{1}}k_{j_{2}}...k_{j_{n}}}$ and
the remaining components $\rho_{k_{j_{1}}},....,\rho_{k_{j_{1}}k_{j_{2}}...k_{j_{n-1}}}$
of the density operator. On the other hand, the second order equations
read
\begin{equation}
\ddot{\rho}_{k_{j_{1}}k_{j_{2}}...k_{j_{n}}}=\overset{N-1}{\underset{k_{j_{1}}...k_{j_{n}},l_{m_{1}}...l_{m_{n}}=1}{\sum}}\mathcal{S}_{k_{j_{1}}k_{j_{2}}...k_{j_{n}}l_{m_{1}}l_{m_{2}}...l_{m_{n}}}\rho_{l_{m_{1}}l_{m_{2}}...l_{m_{n}}}+\mathcal{T}_{k_{j_{1}}k_{j_{2}}...k_{j_{n}}}
\end{equation}
with $\mathcal{S}_{k_{j_{1}}k_{j_{2}}...k_{j_{n}}l_{m_{1}}l_{m_{2}}...l_{m_{n}}}$
and $\mathcal{T}_{k_{j_{1}}k_{j_{2}}...k_{j_{n}}l_{m_{1}}l_{m_{2}}...l_{m_{n}}}$
taking a form comparable to that of the bipartite case. The complicated
form of the dynamics simplifies considerably in the case that $\rho$
is a product state with
\[
\rho_{k_{j_{1}}k_{j_{2}}...k_{j_{n}}}=\alpha^{(1)}_{k_{j_{1}}}\alpha^{(2)}_{k_{j_{2}}}...\alpha^{(n)}_{k_{j_{n}}}
\]
and the interaction parts of the Hamiltonian are zero, since only
then the dynamical equations $(91)$ resp. $(92)$ can be expected
to fully decouple. Whether this is also the case if the separability
criterion $\rho_{k_{j_{1}}k_{j_{2}}...k_{j_{n}}}=\overset{\mu}{\underset{\nu=1}{\sum}}\alpha^{(1)}_{k_{j_{1}}\nu}\alpha^{(2)}_{k_{j_{2}}\nu}...\alpha^{(n)}_{k_{j_{n}}\nu}$
discussed in \cite{de2007separability,hassan2008separability} is
met for all $n$ and the Hamiltonian additionally has a suitable form,
remains an open question. 

Directly solving these equations, as should be noted, remains extremely
difficult. It therefore seems more appropriate to take a different
approach to the subject. One approach is to employ the Dyson series
expansion

\begin{align}
\rho(t) & =\mathcal{T}\exp\left(-i\overset{t}{\underset{0}{\int}}ad_{H(t')}dt'\right)\rho(0)=\rho(0)-i\overset{t}{\underset{0}{\int}}ad_{H(t_{1})}\rho(0)dt_{1}+\\
 & -\overset{t}{\underset{0}{\int}}\overset{t_{1}}{\underset{0}{\int}}ad_{H(t_{1})}ad_{H(t_{1})}\rho(0)dt_{1}dt_{2}+...\nonumber 
\end{align}
with $ad_{H(t)}\cdot=[H(t),\cdot]$, which allows solutions to be
determined directly by calculating the coefficients $\rho_{k_{j_{1}}}:=\frac{d}{2}Tr(\rho\lambda{}^{(j_{1})}_{k_{j_{1}}})$,
$\rho_{k_{j_{1}}k_{j_{2}}}:=\frac{d^{2}}{4}Tr(\rho\lambda{}^{(j_{1}j_{2})}_{k_{j_{1}}k_{j_{2}}})$,...,$\rho_{k_{j_{1}}k_{j_{2}}...k_{j_{n}}}:=\frac{N}{2^{n}}Tr(\rho\lambda{}^{(j_{1}j_{2}...j_{n})}_{k_{j_{1}},k_{j_{2}},...k_{j_{n}}})$.
This approach proves particularly useful in cases where the Hamiltonian
is time-independent and therefore relatively straightforward to diagonalize,
resulting in the expression

\begin{equation}
\rho(t)=\overset{d-1}{\underset{m,n=0}{\sum}}e^{-i(E_{m}-E_{n})t}\rho_{mn}(0)\vert E_{m}\rangle\langle E_{n}\vert
\end{equation}
and associated coefficients $\rho_{k_{j_{1}}}$, $\rho_{k_{j_{1}}k_{j_{2}}}$,
..., $\rho_{k_{j_{1}}k_{j_{2}}...k_{j_{n}}}$. 

To illustrate the effectiveness of this method and thereby conclude
this section, let a specific example be treated in the following.
The choice here falls on the manageable yet illustrative example of
a Heisenberg dimer system that consists of two qubits (spin-one-half
particles) coupled to a small magnetic field. The Hamiltonian of this
model reads
\begin{equation}
H=J\overset{3}{\underset{k=1}{\sum}}(\sigma_{k}\otimes\sigma_{k})+\frac{B}{2}(\mathbf{1}_{2}\otimes\sigma_{3}+\sigma_{3}\otimes\mathbf{1}_{2}),
\end{equation}
where the magnetic field evidently breaks the isotropy of the model.
Taking into account that $h_{j0}=\frac{B}{2}\delta_{j3}$ and $h_{jk}=J\delta_{jk}$
applies in this context and using the Fano-expansion 
\begin{equation}
\rho=\frac{1}{4}\textbf{1}_{2}\otimes\textbf{1}_{2}+\overset{3}{\underset{j=0}{\sum}}r_{j}\sigma_{j}\otimes\textbf{1}_{2}+\overset{3}{\underset{k=0}{\sum}}s_{k}\textbf{1}_{2}\otimes\sigma_{k}+\overset{3}{\underset{j,k=0}{\sum}}t_{jk}\sigma_{j}\otimes\sigma_{k}
\end{equation}
of the density matrix, the results
\begin{align}
\dot{r}_{j} & =\frac{1}{2}(B\overset{3}{\underset{m=0}{\sum}}\epsilon_{3mj}r_{m}+\overset{3}{\underset{m,n=0}{\sum}}J\epsilon_{mnj}t_{mn}),\\
\dot{s}_{k} & =\frac{1}{2}(B\overset{3}{\underset{m=0}{\sum}}\epsilon_{3mk}s_{m}+\overset{3}{\underset{m,n=0}{\sum}}J\epsilon_{nmk}t_{mn}),\nonumber \\
\dot{t}_{jk} & =\frac{J}{2}[\overset{3}{\underset{m=0}{\sum}}\epsilon_{jmk}(r_{m}+s_{m})]+\frac{B}{2}\overset{3}{\underset{m=0}{\sum}}(\epsilon_{3mk}t_{jm}+\epsilon_{3jm}t_{mk})\nonumber 
\end{align}
can be straightforwardly deduced, either by direct calculation or
by using equation $(83)$. The same follows as regards the related
system $(84)$, as can be proven with a little more effort. 

While the resultant system of relations $(83)$ can be decoupled with
some effort, it is considerably more elegant to use $(94)$ and thus
calculate the expressions

\begin{align}
r_{j} & =\overset{1}{\underset{m,n=0}{\sum}}e^{-i(E_{m}-E_{n})t}\rho_{mn}(0)\langle E_{n}\vert\sigma_{j}\otimes\textbf{1}_{2}\vert E_{m}\rangle,\\
s_{k} & =\overset{1}{\underset{m,n=0}{\sum}}e^{-i(E_{m}-E_{n})t}\rho_{mn}(0)\langle E_{n}\vert\textbf{1}_{2}\otimes\sigma_{k}\vert E_{m}\rangle,\nonumber \\
t_{jk} & =\overset{1}{\underset{m,n=0}{\sum}}e^{-i(E_{m}-E_{n})t}\rho_{mn}(0)\langle E_{n}\vert\sigma_{j}\otimes\sigma_{k}\vert E_{m}\rangle;\nonumber 
\end{align}
thereby leading to an exact solution of the von Neumann equation $(1)$,
which, despite the comparatively simple nature of the Hamiltonian,
provides a nontrivial example of the formalism. As can readily be
noted, this method becomes more and more effective the higher the
dimension of the Hilbert space under consideration and the more Bloch
expansion coefficients must therefore be taken into account. 

The Heisenberg dimer, though elementary, is of course not the only
application of equations $(83)$ and $(84)$. Another concrete example
is that of two interacting qubits, i.e., two spin one -half particles,
which are described by a generalized time-dependent Heisenberg model.
Such a model was introduced in \cite{grimaudo2016exactly} and later
discussed in a broader context in \cite{ghiu2020quantum}. A further
example is provided by the models presented in \cite{grimaudo2017spin,grimaudo2019landau},
which deal with coupled qutrits or systems of spin-one half-particles
coupled by higher-order interaction terms \cite{grimaudo2018cooling}.
Furthermore, it is to be expected that there are many more examples,
some of which may be still unknown to the literature. Finding these
examples and coming up with new types of solutions to the equations
mentioned is not the main focus of this paper, though, and will only
be left to future research. Instead, the following concluding section
of this paper discusses two other potential applications of the formalism
presented, thereby showing that said formalism provides concrete physical
predictions that could be tested experimentally or demonstrated in
numerical simulations.

\section{Specific Applications: Intermediate Axis Theorems and Dynamical Entanglement}

Having discussed the above, this section ultimately discusses specific
applications and predictions of the formalism used.

To begin with, based on the validity of $(35)$, let it be noted that
the formalism used leads naturally to a Bloch analog of the intermediate
axis theorem of classical Newtonian mechanics and even to a (higher-dimensional)
generalization thereof. To see this, let the case be considered in
which $\omega_{k}(t)=\omega_{(0)k}+\delta\omega_{k}(t)$ with $\delta\omega_{k}(t)$
small and $\omega_{(0)k}=const.$ applies so that $\dot{\omega}_{k}\approx0$
is satisfied in $(34)$. In the face of this small perturbation around
$\omega_{(0)k}$, a linearization of Euler's equations yields

\begin{equation}
\mathcal{I}_{k}\delta\dot{\omega}_{k}=\overset{3}{\underset{j=1}{\sum}}L_{kj}\delta\omega_{j},
\end{equation}
leading to the characteristic frequencies $\Omega_{kj(m)}=2\sqrt{\alpha_{m}\frac{(\mathcal{I}_{k}-\mathcal{I}_{m})(\mathcal{I}_{j}-\mathcal{I}_{m})}{\mathcal{I}_{k}\mathcal{I}_{j}}}$
with $m$ fixed. To learn about the stability of the system, let it
be assumed that $\mathcal{I}_{3}\leq\mathcal{I}_{2}\leq\mathcal{I}_{1}$
is satisfied. Given this assumption, the following cases can be distinguished:
First of all, there is the case $k=2$, $j=3$, $m=1$ in which $\alpha_{1}=\omega^{2}_{(0)1}$
and thus $\frac{(\mathcal{I}_{1}-\mathcal{I}_{3})(\mathcal{I}_{2}-\mathcal{I}_{1})}{\mathcal{I}_{2}\mathcal{I}_{3}}\leq0$
applies, thereby implying that the rotation around the $\mathcal{I}_{1}$-axis
is stable. Second, there is the case $k=2$, $j=1$, $m=3$ in which,
similarly, $\alpha_{3}=\omega^{2}_{(0)3}$ and thus $\frac{(\mathcal{I}_{1}-\mathcal{I}_{2})(\mathcal{I}_{3}-\mathcal{I}_{2})}{\mathcal{I}_{1}\mathcal{I}_{2}}\leq0$
applies. The motion around the $\mathcal{I}_{3}$-axis is thus stable
as well. However, when considering the cases $k=1$, $j=3$, $m=2$
and $k=3$, $j=1$, $m=2$ in which $\alpha_{3}=\omega^{2}_{(0)2}$
and therefore $\frac{(\mathcal{I}_{3}-\mathcal{I}_{2})(\mathcal{I}_{2}-\mathcal{I}_{1})}{\mathcal{I}_{2}\mathcal{I}_{3}}\geq0$
and $\frac{(\mathcal{I}_{3}-\mathcal{I}_{1})(\mathcal{I}_{3}-\mathcal{I}_{2})}{\mathcal{I}_{1}\mathcal{I}_{3}}\geq0$
is satisfied, one comes to the conclusion that the rotation around
the $\mathcal{I}_{2}$-axis is unstable. As a result, even a slight
perturbation, even for very small initial values $\omega_{(0)1}$
and $\omega_{(0)3}$, causes the object to flip. For the given simplest
possible type of Bloch motion, the same physical behavior can thus
be observed as in the dynamics of rigid bodies in classical Newtonian
mechanics. This shows that the intermediate axis theorem also applies
in full to qubit dynamics and motions of Bloch vectors within the
Bloch sphere.

This raises the question of whether or not the theorem extends to
the higher-dimensional qudit case. To answer this question, it makes
sense to first examine the next higher-dimensional case, namely the
qutrit case. In this case, the generalized Bloch-type Euler equations
$(34)$ take the form

\begin{align}
\mathcal{I}_{1}\dot{\omega}_{1}-(\mathcal{I}_{2}-\mathcal{I}_{3})\omega_{2}\omega_{3}-\frac{1}{2}(\mathcal{I}_{4}-\mathcal{I}_{7})\omega_{4}\omega_{7}-\frac{1}{2}(\mathcal{I}_{5}-\mathcal{I}_{6})\omega_{5}\omega_{6} & =0\\
\mathcal{I}_{2}\dot{\omega}_{2}-(\mathcal{I}_{3}-\mathcal{I}_{1})\omega_{1}\omega_{3}-\frac{1}{2}(\mathcal{I}_{4}-\mathcal{I}_{6})\omega_{4}\omega_{6}-\frac{1}{2}(\mathcal{I}_{5}-\mathcal{I}_{7})\omega_{5}\omega_{7} & =0\nonumber \\
\mathcal{I}_{3}\dot{\omega}_{3}-(\mathcal{I}_{1}-\mathcal{I}_{2})\omega_{1}\omega_{2}-\frac{1}{2}(\mathcal{I}_{4}-\mathcal{I}_{5})\omega_{4}\omega_{5}-\frac{1}{2}(\mathcal{I}_{6}-\mathcal{I}_{7})\omega_{6}\omega_{7} & =0\nonumber \\
\mathcal{I}_{4}\dot{\omega}_{4}-\frac{1}{2}(\mathcal{I}_{7}-\mathcal{I}_{1})\omega_{1}\omega_{7}-\frac{1}{2}(\mathcal{I}_{2}-\mathcal{I}_{6})\omega_{2}\omega_{6}-\frac{1}{2}(\mathcal{I}_{3}-\mathcal{I}_{5})\omega_{3}\omega_{5} & -\frac{\sqrt{3}}{2}(\mathcal{I}_{5}-\mathcal{I}_{8})\omega_{5}\omega_{8}=0\nonumber \\
\mathcal{I}_{5}\dot{\omega}_{5}-\frac{1}{2}(\mathcal{I}_{1}-\mathcal{I}_{6})\omega_{1}\omega_{6}-\frac{1}{2}(\mathcal{I}_{2}-\mathcal{I}_{7})\omega_{2}\omega_{7}-\frac{1}{2}(\mathcal{I}_{3}-\mathcal{I}_{4})\omega_{3}\omega_{4} & -\frac{\sqrt{3}}{2}(\mathcal{I}_{4}-\mathcal{I}_{8})\omega_{4}\omega_{8}=0\nonumber \\
\mathcal{I}_{6}\dot{\omega}_{6}-\frac{1}{2}(\mathcal{I}_{5}-\mathcal{I}_{1})\omega_{1}\omega_{5}-\frac{1}{2}(\mathcal{I}_{2}-\mathcal{I}_{4})\omega_{2}\omega_{4}-\frac{1}{2}(\mathcal{I}_{3}-\mathcal{I}_{7})\omega_{3}\omega_{7} & -\frac{\sqrt{3}}{2}(\mathcal{I}_{7}-\mathcal{I}_{8})\omega_{7}\omega_{8}=0\nonumber \\
\mathcal{I}_{7}\dot{\omega}_{7}-\frac{1}{2}(\mathcal{I}_{1}-\mathcal{I}_{4})\omega_{1}\omega_{4}-\frac{1}{2}(\mathcal{I}_{5}-\mathcal{I}_{2})\omega_{2}\omega_{5}-\frac{1}{2}(\mathcal{I}_{3}-\mathcal{I}_{6})\omega_{3}\omega_{6} & -\frac{\sqrt{3}}{2}(\mathcal{I}_{6}-\mathcal{I}_{8})\omega_{6}\omega_{8}=0\nonumber \\
\mathcal{I}_{8}\dot{\omega}_{8}-\frac{\sqrt{3}}{2}(\mathcal{I}_{5}-\mathcal{I}_{4})\omega_{4}\omega_{5}-\frac{\sqrt{3}}{2}(\mathcal{I}_{7}-\mathcal{I}_{6})\omega_{6}\omega_{7} & =0.\nonumber 
\end{align}
Using the techniques discussed in the previous section, these equations
prove to be integrable, which can also be readily concluded from the
Lax structure of $(33)$. By this latter observation, the relevant
operator-valued constants of motion can be determined rather straightforwardly
from $\mathcal{H}=\frac{1}{2}\overset{8}{\underset{k=1}{\sum}}\ell_{k}\omega_{k}=\frac{1}{2}\overset{8}{\underset{k=1}{\sum}}\mathcal{I}_{k}\omega^{2}_{k},\ell^{2}=\overset{8}{\underset{k=1}{\sum}}\ell^{2}_{k}=\overset{8}{\underset{k=1}{\sum}}\mathcal{I}^{2}_{k}\omega^{2}_{k}$
and $\mathscr{C}{}_{k}=tr\mathfrak{H}^{k}$, resulting in the independent
expressions
\begin{equation}
\mathscr{C}_{1}=\overset{8}{\underset{j,k=1}{\sum}}\delta{}_{jk}\ell_{i}\ell_{j},\mathscr{C}_{2}=\overset{8}{\underset{i,j,k=1}{\sum}}g_{ijk}\ell_{i}\ell_{j}\ell_{k},...
\end{equation}
which are associated with the operator constants of motion $\mathscr{C}_{k}:=\mathcal{C}^{k}$
with $k=1,2,...,N-1$ and $\mathfrak{E}=\{\mathcal{C},\mathfrak{H}\}$.
Moreover, the linear operators
\[
\mathfrak{O}_{1}=\overset{8}{\underset{j,k=1}{\sum}}\delta{}_{jk}\omega_{i}\omega_{j}\textbf{1}_{d},\:\mathfrak{O}_{2}=\overset{8}{\underset{i,j,k=1}{\sum}}g_{ijk}\omega_{i}\omega_{j}\Lambda_{k},...
\]
with $\mathfrak{O}_{k}:=\mathfrak{H}^{k}$ as well as the $SU(3)$
Casimir operators
\begin{equation}
\mathfrak{C}_{1}=\overset{8}{\underset{j,k=1}{\sum}}\delta{}_{jk}\Lambda_{j}\Lambda_{k},\;\mathfrak{C}_{2}=\overset{8}{\underset{i,j,k=1}{\sum}}g_{ijk}\Lambda_{i}\Lambda_{j}\Lambda_{k},...
\end{equation}
 can be identified, which, as follows directly from the Lax pair formulation
of the theory, constitute fully-fledged operator-valued constants
of motion. 

With that clarified, the question arises as to the stability of the
quantum system considered. To address this rather complex question,
it makes sense to linearize the Euler equations around stationary
solutions, which can be accomplished in the given case by considering
the Cartan subalgebra of $SU(3)$ (i.e., the maximal Abelian subalgebra
consisting of diagonal resp. diagonalizable matrices), which can be
constructed from Abelian basis elements $\{\Lambda^{D}_{k}\}$ of
the generalized Gell-Mann basis. The latter is a two-dimensional subalgebra
that can be constructed from the Abelian elements $\{\Lambda_{3},\Lambda_{8}\}$
of the Gell-Mann basis, which give rise to the 'internal axes of symmetry'
of the algebra, i.e. the pair of independent directions $\{e_{3},e_{8}\}$.
These directions allow to specify a constant vector of the form $\omega_{0}=\omega_{(0)3}e_{3}+\omega_{(0)8}e_{8}$,
which gives stationary solutions of the $SU(3)$ Bloch-type Euler
equations $(100)$. Using this fact, the linearized Euler equations
\begin{equation}
\mathcal{I}_{k}\delta\dot{\omega}_{k}=\overset{8}{\underset{j=1}{\sum}}L_{kj}\delta\omega_{j},
\end{equation}
can rather straightforwardly be deduced, provided that the definition
\begin{equation}
L_{kj}=2(f_{3kj}\mathcal{I}_{3}+f_{k3j}\mathcal{I}_{k})\omega_{(0)3}+2(f_{8kj}\mathcal{I}_{8}+f_{k8j}\mathcal{I}_{k})\omega_{(0)8}
\end{equation}
is used in this context. The linearization around a Cartan equilibrium
thus breaks down into three independent $2\times2$-blocks, all of
which must lead to purely imaginary eigenvalues in order to ensure
linear stability of the model. Linearization then ultimately leads
to a single differential equation of the form $(103)$, whereby the
characteristic frequencies read
\begin{equation}
\Omega_{jk(H_{m})}=2\sqrt{\alpha_{H_{m}}\frac{(\mathcal{I}_{j}-\mathcal{I}_{H_{m}})(\mathcal{I}_{k}-\mathcal{I}_{H_{m}})}{\mathcal{I}_{j}\mathcal{I}_{k}}}.
\end{equation}
More specifically, since each factor $\alpha_{H_{m}}$ and each moment
of inertia $\mathcal{I}_{H_{m}}$ varies for each block, three different
types of characteristic frequencies can be deduced in this context,
reading $\Omega_{12(H_{1})}=2\sqrt{\alpha_{H_{1}}\frac{(\mathcal{I}_{1}-\mathcal{I}_{H_{1}})(\mathcal{I}_{2}-\mathcal{I}_{H_{1}})}{\mathcal{I}_{1}\mathcal{I}_{2}}}$
with $\alpha_{H_{1}}\equiv-\omega^{2}_{(0)3}$ and $\mathcal{I}_{H_{1}}\equiv\mathcal{I}_{3}$,
$\Omega_{45(H_{2})}=2\sqrt{\alpha_{H_{2}}\frac{(\mathcal{I}_{4}-\mathcal{I}_{H_{2}})(\mathcal{I}_{5}-\mathcal{I}_{H_{2}})}{\mathcal{I}_{4}\mathcal{I}_{5}}}$
with $\alpha_{H_{2}}\equiv-(\omega_{(0)3}+\sqrt{3}\omega_{(0)8})^{2}$
and $\mathcal{I}_{H_{2}}\equiv\frac{1}{2}(-\mathcal{I}_{3}+\sqrt{3}\mathcal{I}_{8})$
as well as $\Omega_{67(H_{3})}=2\sqrt{\alpha_{H_{3}}\frac{(\mathcal{I}_{6}-\mathcal{I}_{H_{3}})(\mathcal{I}_{7}-\mathcal{I}_{H_{3}})}{\mathcal{I}_{6}\mathcal{I}_{7}}}$
with $\alpha_{H_{3}}\equiv(\omega_{(0)3}+\sqrt{3}\omega_{(0)8})^{2}$
and $\mathcal{I}_{H_{3}}\equiv-\frac{1}{2}(\mathcal{I}_{3}+\sqrt{3}\mathcal{I}_{8})$,
where $\mathcal{I}_{3},\mathcal{I}_{8}<0$ has to be required in the
latter case in order to guarantee that $\mathcal{I}_{H_{3}}$ is meaningful.
Therefore, after assuming that $\mathcal{I}_{8}\leq\mathcal{I}_{7}\leq...\leq\mathcal{I}_{1}$
applies in this context, it becomes clear that necessary and sufficient
conditions for linear stability are $\mathcal{I}_{3}\in(\mathcal{I}_{1,}\mathcal{I}_{2})$
or $\mathcal{I}_{3}\in(\mathcal{I}_{2,}\mathcal{I}_{1})$, $\mathcal{I}_{H_{2}}\in(\mathcal{I}_{4,}\mathcal{I}_{5})$
and $\mathcal{I}_{H_{3}}\in(\mathcal{I}_{6,}\mathcal{I}_{7})$. Moreover,
it becomes apparent that resonance effects occur for $\omega^{2}_{(0)3}=3\omega^{2}_{(0)8}$,
which have no direct counterpart in the more familiar $SU(2)$-setting. 

To prove the complete stability of the system, i.e., linear as well
as nonlinear stability, the Energy-Casimir method described in section
two of this paper can be applied. As a basis for doing this, let the
expression
\begin{equation}
\mathscr{H}=\mathcal{H}+\mu\mathscr{C}_{1}+\nu\mathscr{C}_{2}
\end{equation}
be considered, where $\mathscr{C}_{1}=\overset{8}{\underset{j,k=1}{\sum}}\delta_{jk}\omega_{j}\omega_{k}$
and $\mathscr{C}_{2}=\overset{8}{\underset{i,j,k=1}{\sum}}g_{ijk}\omega_{i}\omega_{j}\omega_{k}$
. In the non-resonant case $\omega^{2}_{(0)3}\neq3\omega^{2}_{(0)8}$,
the method yields 
\begin{equation}
\mu=\frac{-\mathcal{I}_{3}\omega^{2}_{(0)3}+(\mathcal{I}_{3}+2\mathcal{I}_{8})\omega^{2}_{(0)8}}{2(\omega^{2}_{(0)3}-3\omega^{2}_{(0)8})},\;\nu=\frac{(\mathcal{I}_{3}-\mathcal{I}_{8})\omega_{(0)8}}{\sqrt{3}(\omega^{2}_{(0)3}-3\omega^{2}_{(0)8})}
\end{equation}
for the introduced Lagrangian multipliers, thereby making it clear
that the resonance case must be treated separately. Consequently,
the diagonal block terms $h_{j}:=h_{jj}$ (with $j$ fixed) resulting
from the Hessian form read
\begin{equation}
h_{j}=[\mathcal{I}_{j}+2\mu+6\nu\overset{8}{\underset{l=1}{\sum}}g_{jjl}\omega_{l}]\vert_{\omega_{0}}.
\end{equation}
This expression can considerably be simplified under the symmetry
assumptions $\mathcal{I}_{1}=\mathcal{I}_{2}$ and $\mathcal{I}_{4}=\mathcal{I}_{5}=\mathcal{I}_{6}=\mathcal{I}_{7}$,
which yields
\begin{align}
h_{1} & =h_{2}=\frac{\mathcal{I}_{1}-\mathcal{I}_{3}}{2},\\
h_{i} & =\frac{(\mathcal{I}_{i}-\mathcal{I}_{3})\omega^{2}_{(0)3}+\sqrt{3}(\mathcal{I}_{3}-\mathcal{I}_{8})\omega_{(0)3}\omega_{(0)8}+3(\mathcal{I}_{8}-\mathcal{I}_{i})\omega^{2}_{(0)8}}{2(\omega^{2}_{(0)3}-3\omega^{2}_{(0)8})},\nonumber \\
h_{k} & =\frac{(\mathcal{I}_{k}-\mathcal{I}_{3})\omega^{2}_{(0)3}-\sqrt{3}(\mathcal{I}_{3}-\mathcal{I}_{8})\omega_{(0)3}\omega_{(0)8}+3(\mathcal{I}_{8}-\mathcal{I}_{k})\omega^{2}_{(0)8}}{2(\omega^{2}_{(0)3}-3\omega^{2}_{(0)8})}
\end{align}
for $i\in\{4,5\}$ and $k\in\{6,7\}$, where $(107)$ has been used
to eliminate the Lagrange multipliers $\mu$ and $\nu$. That said,
the requirement $h_{1}=h_{2}\geq0$ thus shows that the system is
both linearly stable and Lyapunov stable around the $\mathcal{I}_{1,2}$-axes
in the considered special case. However, additional constraints must
be met to ensure nonlinear stability even under the symmetry assumptions
made, as follows directly from the requirements $h_{4}=h_{5}\geq0$
and $h_{6}=h_{7}\geq0$. Therefore, nonlinear perturbations can cause
runaway behavior even in the mentioned symmetric case, which marks
an important difference from the 'standard' Euler-Poinsot system resulting
in the $SU(2)$-limit. In summary, it can be stated that in the $SU(3)$-case
under consideration, there are three intermediate axis theorems (one
for each root), which already reveals that the generalized Euler-Poinsot
Bloch equations resulting from the qutrit dynamics exhibit a much
richer dynamical structure than the conventional Euler-Poinsot Bloch
equations occurring in the qubit case. Thus, to conclude, let it be
noted that the eight-dimensional $SU(3)$ Bloch geometry has an intrinsic
stability structure that generalizes the conventional intermediate
axis theorem to higher dimensions. Something similar can likewise
be expected to occur in yet higher dimensions, whereby the dynamical
structure is likely to be even richer and thus even more complex. 

Having clarified all this, it may now be time to discuss another application
of the formalism used. This application focuses on the construction
of specific solutions of the Bloch vector equations $(13)$, which
shall be referred to as dynamically entangled states in the following.
These states change their entanglement structure by design over time
from separable to maximally entangled and vice versa. To give a simple,
but yet nontrivial example for this, let a bipartite qubit Hilbert
space $\mathcal{H}\equiv\mathcal{H}_{1}\otimes\mathcal{H}_{2}$ and
an associated quantum system be considered, which has the special
property that the Bloch vector $\vec{\omega}$ of the Hamiltonian
and the Bloch vector $\vec{\ell}$ of the observable $\mathcal{C}$
depicted in $(20)$ are both constant. A simple example of a Hamiltonian
that meets exactly this condition is 
\begin{equation}
H=\frac{\omega}{2}(\sigma_{1}\otimes\sigma_{1}-\sigma_{2}\otimes\sigma_{2})=\omega(\vert00\rangle\langle11\vert+\vert11\rangle\langle00\vert),
\end{equation}
as can readily be checked by expanding the solution in an $SU(4)$
generalized Gell-Mann basis $\{\Lambda_{1},\Lambda_{2},....,\Lambda_{15}\}$,
which shows that the only non-zero components are $\omega_{9}=\ell_{9}=\omega=const.$
The unique time-dependent solution of the von Neumann equation $(1)$
corresponding to this particular type of Hamiltonian is 
\begin{align}
\rho(t) & =\frac{1}{4}(\textbf{1}_{2}\otimes\textbf{1}_{2}+\cos2\omega t(\sigma_{3}\otimes\textbf{1}_{2}+\textbf{1}_{2}\otimes\sigma_{3})+\sigma_{3}\otimes\sigma_{3}+\\
 & +\sin2\omega t(\sigma_{1}\otimes\sigma_{1}-\sigma_{2}\otimes\sigma_{2})),\nonumber 
\end{align}
that is, a solution $\rho(t)=\vert\psi(t)\rangle\langle\psi(t)\vert$
that can be written as a time-dependent pure state
\begin{equation}
\vert\psi(t)\rangle=\cos\omega t\vert00\rangle+\sin\omega t\vert11\rangle.
\end{equation}
As can readily be checked, this state is separable for $t=\frac{n}{4}T$
and maximally entangled for $t=\frac{n}{8}T$. Thus, one comes to
the conclusion that the state performs what shall be referred to as
an entanglement oscillation, where the state changes its entanglement
structure in the time period $t\in[\frac{n}{8}T,\frac{n}{4}T]$ for
$n\in\mathbb{N}_{0}$. 

A natural higher-dimensional generalization of the above arises in
case that a bipartite qudit Hilbert space $\mathcal{H}\equiv\mathcal{H}_{1}\otimes\mathcal{H}_{2}$
is considered. In this case, by analogy, the constant Hamiltonian
\begin{equation}
H=\omega(\vert00\rangle\langle d-1d-1\vert+\vert d-1d-1\rangle\langle00\vert)
\end{equation}
can be considered, which then yields a pure state of the form
\begin{equation}
\vert\psi(t)\rangle=\cos\omega t\vert00\rangle+\sin\omega t\vert d-1d-1\rangle,
\end{equation}
as follows again directly from the von Neumann equation $(1)$. This
state again changes its entanglement structure in the time period
$t\in[\frac{n}{8}T,\frac{n}{4}T]$, thereby giving again an example
for the effect of entanglement oscillation. 

Ultimately, a further generalization can be made in this context by
transitioning to a multipartite qudit state of the form 
\begin{equation}
\vert\psi(t)\rangle=\cos\omega t\vert kk...k\rangle+\sin\omega t\vert\phi(t)\rangle,
\end{equation}
with
\begin{equation}
\vert\phi(t)\rangle=\overset{d-1}{\underset{\underset{j\neq k}{j=0}}{\sum}}\beta_{j}(t)\vert jj...j\rangle
\end{equation}
and $k\in\{0,1,...,d-1\}$ fixed, where the coefficients $\beta_{j}(t)$
with $\beta_{0}(t)=\overset{d-1}{\underset{m=0}{\prod}}\cos\omega_{m}t$,
$\beta_{1}(t)=\sin\omega_{0}t\left(\overset{d-2}{\underset{m=1}{\prod}}\cos\omega_{m}t\right)$,
..., $\beta_{d-2}(t)=\sin\omega_{d-2}t\cos\omega_{d-1}t$ , $\beta_{d-1}(t)=\sin\omega_{d-1}t$
are the coefficients of a sequence of Givens rotations. Using here
the fact that $\vert\beta_{j}(n\frac{\pi}{4})\vert=\frac{1}{\sqrt{d}}$
for $j=1,2,..,d$ and $\vert\beta_{d-1}(n\frac{\pi}{2})\vert=1$ and
$\vert\beta_{j}(n\frac{\pi}{2})\vert=0$ with $j<d-1$ for $\omega_{1}=\omega_{2}=...=\omega_{d-1}$,
it becomes clear that the entanglement structure of the state changes
also in the given case over time, where the state transitions from
maximally entangled to separable in the time period $t\in[\frac{n}{8}T,\frac{n}{4}T]$.
This state therefore represents an oscillating pure entanglement state
that periodically changes its entanglement structure over time while
remaining pure throughout. Such states arise quite naturally as solutions
to the derived Bloch vector equations, thus providing an example of
how the formalism leads to non-trivial physical applications. It can
be expected that there are further such applications. Finding the
latter, however, is left to future research.

\section*{Conclusion and Outlook}

In this present work, the dynamics of closed quantum systems were
investigated in the Bloch vector representation using methods from
rigid body dynamics and the theory of integrable systems. In doing
so, Bloch vector equations of motion were deduced that correspond
to the Euler-Poinsot equations for torque-free spinning tops in the
qubit case and hitherto unknown generalized Euler-Poinsot-type equations
in the higher-dimensional case. To prove the Liouville integrability
of the corresponding Hamiltonian equations of motion, connections
to the Neumann model of classical Hamiltonian dynamics and to the
Hamilton-Euler-Poinsot model were established in the $SU(2)$-case
in order to identify the first integrals of motion. This method was
generalized using Adler-Kostant-Symes scheme and a Gaudin-type Lax
matrix ansatz to prove integrability of the generic $SU(d)$ Bloch
vector dynamics, whereby a previously unknown explicit closed-form
solution of the Bloch vector dynamics in the form of theta functions
was derived based on the Arnold-Liouville theorem \cite{arnol2013mathematical,babelon2003introduction,landau1982mechanics}.

In this context, quantum analogues of the Routh-Hurwitz criterion
and the Energy-Casimir method were formulated (though only the latter
proves relevant for ensuring the stability of unitary dynamics), novel
types of of Bloch vector equations were derived for multipartite Hilbert
spaces and specific solutions to the corresponding equations of motion
were constructed that encode the complex dynamics of composite quantum
systems. Eventually, two previously unknown physical effects were
predicted: First, a Bloch-analog of the intermediate axis theorem
was derived. Second, toward the end of the work, the effect of oscillating
entanglement was predicted and discussed as a concrete application
of the formalism used. As a basis for the latter, special types of
solutions to the equations of motion were derived that constitute
oscillating entangled states, that is, dynamical quantum states that
change their entanglement structure from maximally entangled to separable
and vice versa. It is to be expected, though, that many more solutions
to the von Neumann and Schrödinger equations can be found that lie
in the constructed integrable $SU(d)$ family.

The formalism used, as should ultimately be noted, still lacks a generalization
to include open quantum systems that are subject to decoherence and
environmental noise. The latter could, however, be achieved quite
simply by turning to the Lindblad case, i.e., by expanding the Lindblad
equation instead of the von Neumann equation in a generalized Gell-Mann
matrix basis and examining the resultant equations of motion for the
Bloch components. This step was deliberately avoided in this study,
however, in order to first gain a better understanding of the simpler
case of closed quantum systems. By focusing exclusively on that simpler
case, though, some questions remained open. These include e.g. whether
the integrability of the derived Bloch vector equations of motion
can also be proven in the Lindblad case using methods similar to those
used in this work, and whether the stability criteria derived for
closed systems can also be applied in the more general context of
open quantum systems. Another question that arises in this context
is whether the resulting stability criteria could be used in applications
in quantum control theory or quantum state tomography, for example,
to design quantum systems that are more controllable in practice and
thus exhibit greater overall robustness than those currently available.
A related open question is whether the findings on the stability of
quantum dynamics could ultimately also be used for quantum error suppression
in quantum computing and related applications. 

Clarifying all these questions in full detail, however, must be deferred
to future work on the subject.
\begin{description}
\item [{Acknowledgements:}]~
\end{description}
Great thanks to Franz Embacher for critical discussions on the subject
and helpful comments on the manuscript. 

\appendix

\section*{Appendix A:}

This part of the paper summarizes a few formal steps, which are not
explicitly shown in the main text of the paper. To begin with, it
may be noted that

\begin{equation}
\frac{d}{dt}\text{Tr}\left(\mathcal{C}^{k-j}K^{j}\right)=(k-j)\text{Tr}\left([K,\mathcal{C}]K^{j}\mathcal{C}^{k-j}\right)=\text{Tr}\left([K,\mathcal{C}]X\right)=0
\end{equation}
holds, where the cyclicity and linearity of the trace have been used
in the final step. Using the definition 
\begin{equation}
\{F,G\}=\underset{i,j,k=1}{\overset{N-1}{\sum}}f_{ijk}\ell_{k}\frac{\partial F}{\partial\ell_{i}}\frac{\partial G}{\partial\ell_{j}},
\end{equation}
it is straightforward to conclude that 
\begin{equation}
\{\mathcal{C},\mathcal{C}\}=\underset{i,j,k=1}{\overset{N-1}{\sum}}f_{ijk}\ell_{k}\frac{\partial\mathcal{C}}{\partial\ell_{i}}\frac{\partial\mathcal{C}}{\partial\ell_{j}}=\underset{i,j,k=1}{\overset{N-1}{\sum}}f_{ijk}\ell_{k}\Lambda_{i}\Lambda_{j}.
\end{equation}
Using the commutator relation depicted in $(4)$, one thus obtains
\begin{align}
\underset{i,j,k=1}{\overset{N-1}{\sum}}f_{ijk}\ell_{k}(\Lambda_{i})_{mn}(\Lambda_{j})_{rs} & =\frac{1}{2i}(\mathcal{C}_{ms}\delta_{nr}-\mathcal{C}_{rn}\delta_{sm})\\
= & ([r,\mathcal{C}\otimes\textbf{1}_{d}+\textbf{1}_{d}\otimes\mathcal{C}])_{mn,rs}\nonumber 
\end{align}
where the object

\begin{equation}
r_{ij,kl}=\frac{1}{2i}\delta_{il}\delta_{jk}
\end{equation}
defines the $r$-matrix. This leads to the result

\begin{equation}
\{\mathcal{C},\mathcal{C}\}=[r,\mathcal{C}\otimes\textbf{1}_{d}+\textbf{1}_{d}\otimes\mathcal{C}],
\end{equation}
whereby the consistency of the result can be checked using the Sklyanin
bracket \cite{babelon2003introduction}
\begin{align}
\{L(\lambda)\overset{\otimes}{,}L(\mu)\} & =\left[\frac{\mathcal{P}}{\lambda-\mu},L(\lambda)\otimes\textbf{1}_{d}+\textbf{1}_{d}\otimes L(\mu)\right]\\
= & \left[\frac{\mathcal{P}}{\lambda-\mu},(\mathcal{C}+\lambda K)\otimes\textbf{1}_{d}+\textbf{1}_{d}\otimes(\mathcal{C}+\mu K)\right]\nonumber \\
= & \left[\frac{\mathcal{P}}{\lambda-\mu},\mathcal{C}\otimes\textbf{1}_{d}+\textbf{1}_{d}\otimes\mathcal{C}\right]+\left[\frac{\mathcal{P}}{\lambda-\mu},\lambda K\otimes\textbf{1}_{d}+\mu\textbf{1}_{d}\otimes K\right].\nonumber 
\end{align}
 The second term vanishes if and only if the condition
\begin{equation}
[\mathcal{P},K\otimes\textbf{1}_{d}+\textbf{1}_{d}\otimes K]=0
\end{equation}
is met, which is tantamount to requiring that $K$ is symmetric. However,
since $K$ is just a multiple of the Hamiltonian operator $H$, it
is by definition a self-adjoint operator. Assuming that the Bloch
vector $\vec{h}$ occurring in relation $(3)$ is real-valued, it
thus becomes clear that $K$ is indeed symmetric and that condition
$(125)$ is fulfilled. But this shows that equations $(124)$ and
$(125)$ are equivalent. 

As should be noted, the objects $r_{12}=r\otimes\textbf{1}_{d}$,
$r_{13}=(\textbf{1}_{d}\otimes\Pi)(r\otimes\textbf{1}_{d})(\textbf{1}_{d}\otimes\Pi)$
and $r_{23}=\textbf{1}_{d}\otimes r$ meet the classical Yang-Baxter
equation

\begin{equation}
[r_{12},r_{13}]+[r_{12},r_{23}]+[r_{13},r_{23}]=0
\end{equation}
and obey the braiding relations
\begin{equation}
\sigma_{i}\sigma_{i+1}\sigma_{i}=\sigma_{i+1}\sigma_{i}\sigma_{i+1},
\end{equation}
where $\sigma_{i}:=\mathcal{P}_{i,i+1}$ applies by definition. See
here e.g. \cite{babelon2003introduction} for further details. 

Ultimately, to prove relation $(76)$, let it be assumed that $\Sigma$
is a smooth, compact Riemann surface of genus $g$, $\theta(u)$ the
corresponding theta function, $\mathcal{A}:\Sigma\rightarrow\text{Jac}(\Sigma)$
the Abel-Jacobi map, and $\vec{u}(t)=\vec{\Omega}t+\vec{u}_{0}$ a
linear flow on the Jacobian. Then the following applies: Any meromorphic
function $f(P,t)$ on $\Sigma$ whose pole divisor is determined by
a finite number of points $P_{i}$ can be expressed locally in the
form
\begin{equation}
f(P,t)=c(P)+\vec{\Omega}(P)\vec{\nabla}_{u}\ln\frac{\Theta(\mathcal{A}(P)+u(t)-\Delta)}{\Theta(u(t)-\Delta)}.
\end{equation}
To see this, consider a system of points $P,Q,R,S\in\Sigma$. Fay's
trisecant identity \cite{fay2006theta} states for such a system of
points
\begin{align}
\Theta & (u+\mathcal{A}(P)-\mathcal{A}(Q))\Theta(u+\mathcal{A}(R)-\mathcal{A}(S))\\
-\Theta & (u+\mathcal{A}(P)-\mathcal{A}(S))\Theta(u+\mathcal{A}(R)-\mathcal{A}(Q))\nonumber \\
= & \Theta(u)\Theta(u+\mathcal{A}(P)+\mathcal{A}(R)-\mathcal{A}(Q)-\mathcal{A}(S))F(P,Q,R,S),\nonumber 
\end{align}
where $F(P,Q,R,S)$ is a function that is independent of $u$. For
$R=P+\epsilon$ and $S=Q+\epsilon$, it then follows \cite{fay2006theta}
in the limit $\epsilon\rightarrow0$
\begin{equation}
\partial_{u_{k}}\ln\frac{\Theta(u+\mathcal{A}(P)-\mathcal{A}(Q))}{\Theta(u)}=\stackrel[Q]{P}{\int}\omega_{k}+\text{reg}.
\end{equation}
Given that, it follows from the theory of Riemann surfaces (Riemann-Roch
theorem \cite{atiyah1959riemann}) that meromorphic functions with
a given divisor are uniquely determined by their pole structure and
can be uniquely represented by theta functions. Since the right-hand
side describes exactly one function with a simple pole at $P$ and
a zero at $Q$, it follows that logarithmic derivatives of theta-quotients
generate exactly these functions. In the present case, the divisor
$\mathcal{D}(t)$ is linear on the Jacobian, i.e. $\vec{u}(t)=\vec{u}_{0}+\vec{\Omega}t$,
thereby implying that $\frac{d}{dt}=\vec{\Omega}\vec{\nabla}_{u}$.
This implies further
\begin{equation}
f(t)=\vec{\Omega}\vec{\nabla}_{u}\ln\frac{\Theta(u+c_{1})}{\Theta(u+c_{2})}
\end{equation}
and so leads to $(128)$. Using the degenerate form of Fay\textquoteright s
identity, logarithmic derivatives of theta-function ratios generate
meromorphic functions on the spectral curve with prescribed divisor.
Since the observables $\ell_{k}(t)$ can be expressed as functions
of the separated variables $(\lambda_{j}(t),\mu_{j}(t))$ they inherit
this structure. Consequently, they admit a representation in terms
of logarithmic derivatives of theta functions, yielding $(76)$.

Note that a concise notation table listing relevant symbols and terms
used in the paper is provided in $\text{Figure 1}$ of the manuscript.

\appendix

\section*{Appendix B:}

The von Neumann equation leads to a relation $(83)$ for Bloch decompositions
of the form $(81)$. The coefficients are
\begin{align}
\mathcal{L}^{(1)}_{il} & :=2\overset{N-1}{\underset{k=1}{\sum}}\mathrm{f}_{ilk}h_{0k},\mathcal{L}^{(2)}_{il}:=2\overset{N-1}{\underset{k=1}{\sum}}\mathrm{f}_{ilk}h_{k0},\mathcal{M}^{(1)}_{i}:=\overset{N-1}{\underset{j,k,m=1}{\sum}}\mathrm{f}_{jki}h_{jm}\rho_{km},\\
\mathcal{M}^{(2)}_{i} & :=\overset{N-1}{\underset{j,k,m=1}{\sum}}\mathrm{f}_{jki}h_{mj}\rho_{km},\;\mathcal{N}_{jklm}:=2\overset{N-1}{\underset{n=1}{\sum}}\mathrm{\mathrm{f}}_{nlj}\delta_{mk}h_{n0}+2\overset{N-1}{\underset{n=1}{\sum}}\mathrm{f}_{nlk}\delta_{mj}h_{0n}+\nonumber \\
\overset{N-1}{\underset{n,o=1}{\sum}} & (\mathrm{f}_{nlj}\mathrm{g}_{omk}+\mathrm{g}_{nlj}\mathrm{f}_{omk})h_{no},\;\mathcal{O}_{jk}:=2\overset{N-1}{\underset{l,m=1}{\sum}}\mathrm{f}_{lmj}\rho_{0l}h_{mk}+2\overset{N-1}{\underset{l,m=1}{\sum}}\mathrm{f}_{lmk}\rho_{l0}h_{mj}.\nonumber 
\end{align}
This gives 
\begin{equation}
\mathcal{Q}^{(j)}_{im}=\dot{\mathcal{L}}^{(j)}_{im}+\overset{N-1}{\underset{l=1}{\sum}}\mathcal{L}^{(j)}_{il}\mathcal{L}^{(j)}_{lm},\mathcal{R}^{(j)}_{i}=\dot{\mathcal{M}}^{(j)}_{i}+\overset{N-1}{\underset{l=1}{\sum}}\mathcal{L}^{(j)}_{il}\mathcal{M}^{(j)}_{l}
\end{equation}
with $j=1,2$ and
\begin{equation}
\mathcal{S}_{jklm}=\mathcal{\dot{N}}_{jklm}+\overset{N-1}{\underset{n,o=1}{\sum}}\mathcal{N}_{jkno}\mathcal{N}_{nolm},\;\mathcal{T}_{jk}=\dot{\mathcal{O}}_{jk}+\overset{N-1}{\underset{l,m=1}{\sum}}\mathcal{N}_{jklm}\mathcal{O}_{lm}
\end{equation}
in $(84)$. Ultimately, the coefficients occurring in $(86)$ can
be determined to be
\begin{align}
\mathcal{U}^{(1)}_{i} & :=\overset{N-1}{\underset{j,k,m=1}{\sum}}\mathrm{f}_{jki}h_{jm}\ell_{km},\mathcal{U}^{(2)}_{i}:=\overset{N-1}{\underset{j,k,m=1}{\sum}}\mathrm{f}_{jki}h_{mj}\ell_{km},\\
\mathcal{V}_{jk} & =2\overset{N-1}{\underset{l,m=1}{\sum}}\mathrm{f}_{lmj}\ell_{0l}h_{mk}+2\overset{N-1}{\underset{l,m=1}{\sum}}\mathrm{f}_{lmk}\ell_{l0}h_{mj}.\nonumber 
\end{align}
The Bloch coefficients occurring in relations $(91)$ and $(92)$
take an analogous, but a more complicated form. 

\bibliographystyle{plain}
\addcontentsline{toc}{section}{\refname}\bibliography{litkrei}

\end{document}